\documentclass[12pt,titlepage]{article}
\usepackage{aliascnt}
\usepackage{amscd}
\usepackage{amsfonts}
\usepackage{amsmath}
\usepackage{amssymb}
\usepackage{amsthm}
\usepackage{array}
\usepackage{bbm}
\usepackage[english]{babel}
\usepackage{dsfont}
\usepackage{enumerate}
\usepackage{graphicx}
\usepackage{caption}
\usepackage{subcaption}
\usepackage{hyperref}
\usepackage{latexsym}
\usepackage{multirow}
\usepackage{epstopdf}
\usepackage{mathrsfs}
\usepackage{natbib}
\usepackage{pdfsync}
\usepackage{pgfplots}
\usepackage{epstopdf}
\usepackage{tikz}
\usepackage[linesnumbered,ruled]{algorithm2e}
\usepackage{algpseudocode}
\usepackage{color}
\usepackage{graphicx,psfrag,epsf}
\usepackage{enumerate}
\usepackage{url}
\usepackage{amsthm,bm}
\renewenvironment{proof}{{\em Proof.} }{\hfill $\Box$}

\newtheorem{theorem}{Theorem}
\newtheorem{definition}{Definition}

\newtheorem{corollary}{Corollary}

\def\1{\mathds{1}}

\usepackage{geometry}
\begin{document}
\pagenumbering{roman}
\title{Contour Models for Boundaries Enclosing Star-Shaped and Approximately Star-Shaped Polygons \thanks{We thank Cecilia Bitz, Donald Percival,  and Daniel Pollack for helpful discussions. Results are generated with the \textit{ContouR} R package \citep{Director2020} and scripts accessible at \url{https://github.com/hdirector/contourPaperScripts}.  Contributions by HMD and AER were supported by NOAA's Climate Program Office, Climate Variability and Predictability Program through grant NA15OAR4310161. HMD's contribution to this work was also based upon work supported by the National Science Foundation Graduate Research Fellowship under Grant No. DGE-1256082. Any opinion, findings, and conclusions or recommendations expressed in this material are those of the authors and do not necessarily reflect the views of the National Science Foundation. 
} }
\author{Hannah M. Director\\ Department of Statistics \\ University of Washington \and
Adrian E. Raftery \\ Departments of Statistics and Sociology \\University of Washington}
\date{
\mbox{} \\
July 8, 2020}
\maketitle 

\begin{abstract}
Boundaries on spatial fields divide regions with particular features from surrounding background areas. These boundaries are often described with contour lines. To measure and record these boundaries, contours are often represented as ordered sequences of spatial points that connect to form a line.  Methods to identify boundary lines from interpolated spatial fields are well-established.  Less attention has been paid to how to model sequences of connected spatial points. For data of the latter form, we introduce the Gaussian Star-shaped Contour Model (GSCM). GSMCs generate sequences of spatial points  via generating sets of distances in various directions from a fixed starting point. The GSCM is designed for modeling contours that enclose regions that are star-shaped polygons or approximately star-shaped polygons. Metrics are introduced to assess the extent to which a polygon deviates from star-shaped.  Simulation studies illustrate the performance of the GSCM in various scenarios and an analysis of Arctic sea ice edge contour data highlights how GSCMs can be applied to observational data.
\end{abstract}

\begin{small}
\newpage
\tableofcontents

%\medskip
\newpage

\listoftables

\medskip

\listoffigures
\end{small}

\newpage
\pagenumbering{arabic}
\baselineskip=18pt

\section{Introduction}
Boundaries that enclose regions are often subjects of scientific interest.  Contour lines divide a contiguous region with some defining feature(s) from surrounding background areas.   In this paper, we focus on how to infer the distribution of contours from multiple fully-observed contours. We assume that observations of contours are sets of ordered, connected spatial points in the 2-D plane.  Sets of connected spatial points can also come from grids of binary values that indicate if each grid box is inside or outside the boundary. The contours in this case are the points that connect to form the boundary between the grid boxes inside and outside the region. Distributions of contours are inferred from observing multiple contours, such as contours observed at different times in a stationary process. We introduce the Gaussian Star-shaped Contour Model (GSCM) for inferring distributions of contours.  GSCMs are designed for modeling contours that enclose regions that are \textit{star-shaped polygons}  (Definition \ref{def: starShapedPolygon}) or approximately star-shaped polygons

\begin{figure}[h!]
\centering
\includegraphics[width=.5\textwidth]{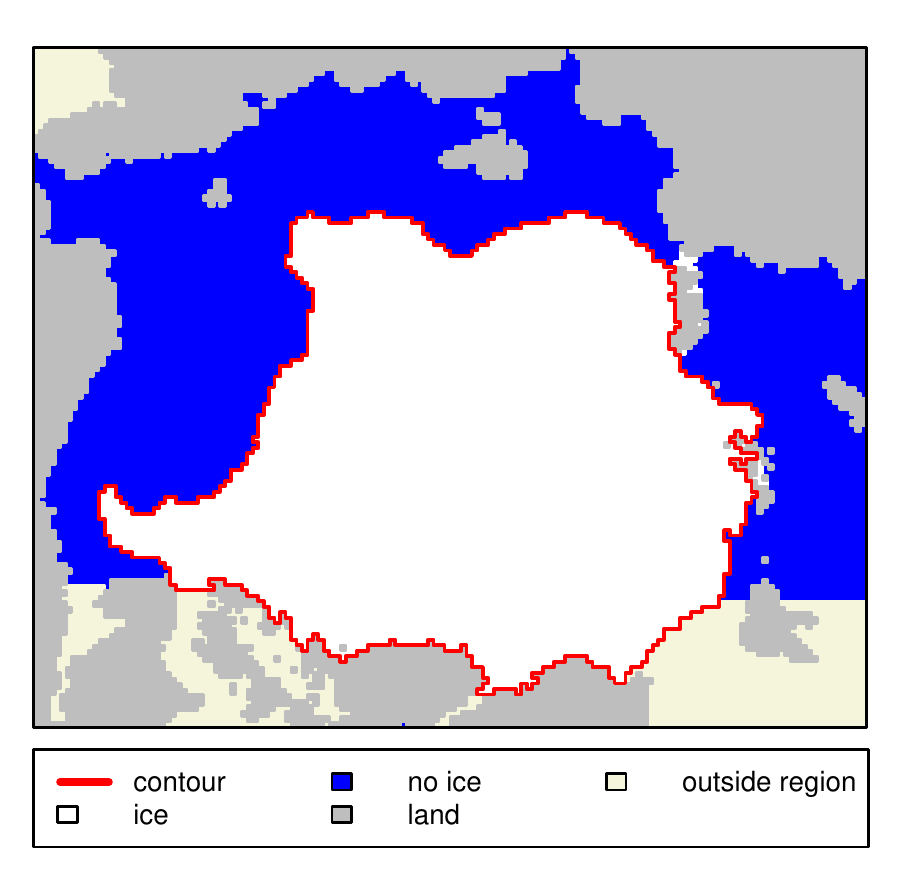}
\caption{The contour forming the boundary around the main contiguous area covered by sea ice in a central region of the Arctic in September 2017. Section \ref{sec: approxStar} introduces methods for modeling contours like this one that enclose approximately star-shaped polygons. } 
\label{fig: sampleIceEdge}
\end{figure}

We consider the Arctic sea ice edge contour as a motivating example. Figure \ref{fig: sampleIceEdge} shows a sample sea ice edge contour. The sea ice edge contour forms a boundary between the area covered by sea ice and the surrounding open water. For sea ice edge contours, questions of both inference and prediction are relevant. Polar scientists are interested in where ice edge contours are more and less variable and the extent to which ice edge contours change over time. Predictions of the ice edge contour are also needed weeks to months in advance for maritime planning.   

Previous research has developed contour and boundary models for other data types. Analysis has often focused on inferring a single boundary from observations on a spatial field. Research on exceedance levels has developed methods to infer contours that describe where a property goes above some level \citep{Bolin2015, French2013, French2016}. Wombling methods find contours by identifying curvilinear gradients \citep{Womble1951, Banerjee2006}. Statistical shape analysis \citep{Dryden2016, Srivastava2016} provides tools to model boundaries corresponding to particular objects with discernible features.  While important statistical techniques, none of these methods infer distributions of contours from multiple observed contours. The GSCM seeks to fill this methodological gap.  

In the remainder of the paper, we develop GSCMs and assess their performance.  Section \ref{sec: probModel}  defines contours and how to represent them. Section \ref{sec: GSCM} introduces the GSCM for modeling contours enclosing star-shaped polygons and discusses model fitting.  Section \ref{sec: coverageMetric} introduces a metric for assessing the extent to which a contour differs from star-shaped and Section \ref{sec: sims} presents simulation studies. Section \ref{sec: approxStar} extends GSCMs to contours enclosing approximately star-shaped polygons and Section \ref{sec: example}  considers an example with sea ice edge contours. Section \ref{sec: discussion} concludes the paper with discussion,  including a more involved discussion of how GSCMs compare to other contour and boundary methods.

\section{Contour definitions}
\label{sec: probModel}
Our focus is on modeling contours that act as the boundary between a region that has some feature(s) and the surrounding background region.  There are multiple  ways  such contours  could be defined. In this section, we give two representations for these contours  that will be used as a basis for subsequent modeling and assessment.

\subsection{Point-sequence representation}
\label{sec: pointSequence}
Contours and the regions they enclose can be described using connected sequences of points. We refer to this description of a contour as a {\it{point-sequence representation}}. We define the following concepts.

 \begin{definition}{Contour point sequence, $\boldsymbol{S}$:}
An ordered set of spatial points $\boldsymbol{S} = (\boldsymbol{s}_{1}, \hdots, \boldsymbol{s}_{n})$, with $n > 2$, where each $\boldsymbol{s}_{i}$ consists of the x-y coordinates of a spatial location.  \end{definition}

\begin{definition}{Contour line, $\boldsymbol{\overline{S}}$:} 
The connected line formed by connecting $\boldsymbol{s}_{i}$ to $\boldsymbol{s}_{i + 1}$ for $i = 1, \hdots, n - 1$ and connecting $\boldsymbol{s}_{n}$ to $\boldsymbol{s}_{1}$.
\end{definition}

\begin{figure}[h!]
\centering
\includegraphics[width=.7\textwidth]{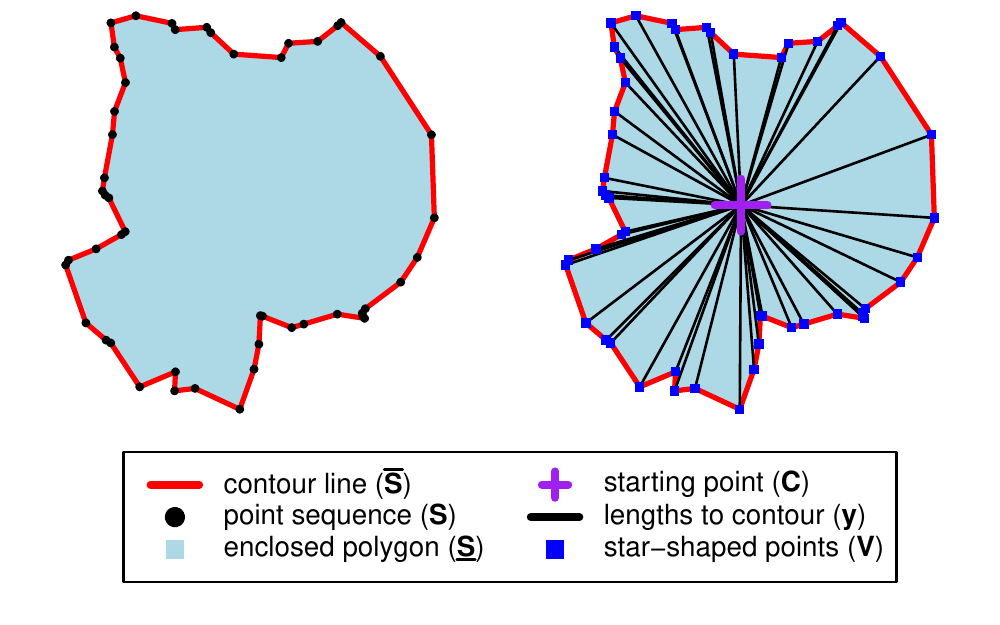}
\caption{Components of a contour represented by a point-sequence representation (left) and a star-shaped representation (right).} 
\label{fig: starVPoint}
\end{figure}

\begin{definition}{Enclosed polygon, $\boldsymbol{\underline{S}}$:} 
The polygon formed by the interior of the contour  line $\boldsymbol{\overline{S}}$.
\end{definition}
The left panel of Figure \ref{fig: starVPoint}  illustrates these definitions for a contour described by a  point-sequence representation.  The main advantage of the point-sequence representation is its flexibility. Any contour enclosing a polygon can be represented exactly with a sequence of spatial points, $\boldsymbol{S}$. Also, the level of detail represented can be increased simply by increasing the number of points.

A binary grid that indicates whether each grid box is inside or outside the region of interest  can also be converted to the point-sequence representation. The contour  $\boldsymbol{S}$ is made up of corner points of grid boxes that touch the outside of the region on one side and the inside of the region on the other. The points of $\boldsymbol{S}$ are ordered to align with the order in which they would be touched if one were to trace around the boundary.   Where and in what direction to start tracing around the boundary is arbitrary. These choices only determine the indexing of the points in $\boldsymbol{S}$,  not the line, $\boldsymbol{\overline{S}}$,  or enclosed polygon, $\boldsymbol{\underline{S}}$. The point-sequence representation can be used with any grid resolution, though  finer grids will require more points in $\boldsymbol{S}$.

\subsubsection{Notation}
We need to distinguish between points, lines, and polygons. An ordered sequence of spatial points will be denoted by a boldface letter, such as $\boldsymbol{S}$. A line formed by connecting points will be denoted with an overline $\overline{\boldsymbol{S}}$.  A line segment will be denoted by an overline over two letters that represent the start and end points of the segment, such as $\overline{\boldsymbol{C}\boldsymbol{D}}$.
The polygon enclosed by a line, $\underline{\boldsymbol{S}}$, will be denoted by an underline.

\subsubsection{Contours with fractal characteristics}
We acknowledge that the point-sequence representation does not directly account for contours whose true nature is fractal. As such, representing fractal contours as connected sequences of points may be an approximation. In contours of sea ice and other physical world examples, as the spatial scale of observations increases, the level of detail of the contour also increases  \citep{Mandelbrot1967}. With each increase in spatial resolution additional line segments are needed to describe the increased detail. In other words, the length of the contour increases each time the spatial resolution increases. This fractal nature of some contours means that these contour can never be fully expressed with a finite, ordered set of spatial points. These contours' true lengths are infinite.

For statistical applications, however, limits exist on the precision of measurements and the relevant scale of scientific interest. While the fractal or Hausdorff dimension can be estimated \citep{Gneiting2012}, the level of detail of the boundary that will be measured or needed will rarely be a fractal. So, making the simplifying assumption that the contour can be defined by a sequence of spatial points is reasonable for many applications.  Additional discussion of contours as fractals is given in Section \ref{sec: fracGSCMs}.

\subsection{Star-shaped representation}
\label{sec: starShapedApprox}
Point-sequence representations are natural and describe contours accurately.  However, point-sequence representations are ill-suited for describing multiple contours and their distributional behavior. Contours differ in length, so two points with the same index on two different  point-sequence representations are not likely to be in the same physical location. Comparing spatially-dependent features and inferring distributions is therefore difficult.  We build an alternate  {\it{star-shaped representation}} that avoids the weaknesses of point-sequence representations.  This representation is appropriate for contours that enclose  star-shaped polygons or approximately star-shaped polygons.

\begin{figure}
\centering
 \includegraphics[clip,width=.5\columnwidth]{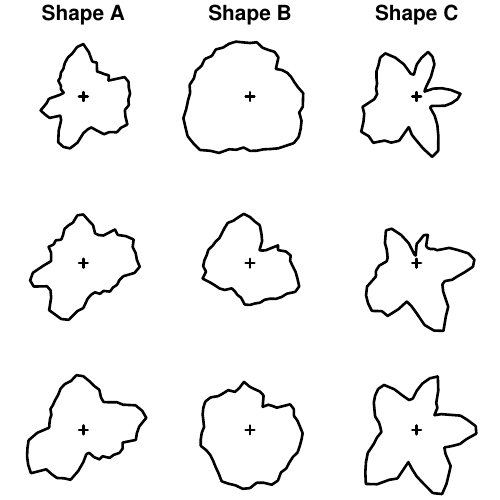}
\centering
\includegraphics[clip,width=.5\columnwidth]{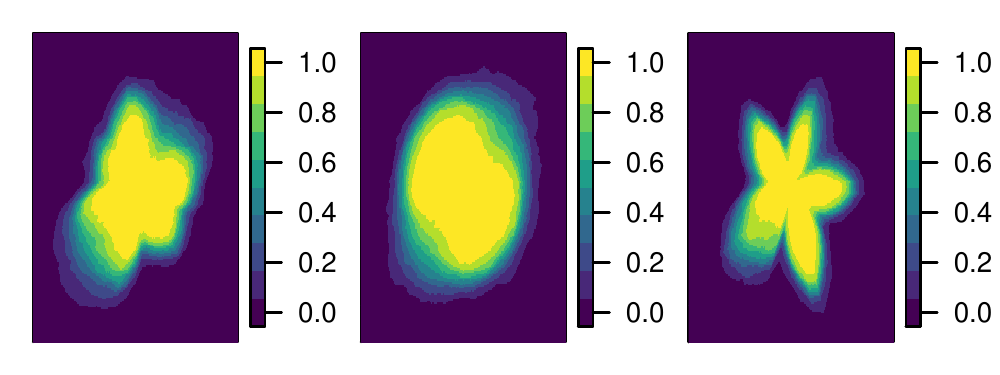}
\caption{Rows 1-3: Nine examples of contours enclosing star-shaped polygons generated from GSCMs with three different parameter settings, organized by column. The cross sign denotes the starting point, $\boldsymbol{C}$. Row 4: Estimated probability of a grid box being contained within contours generated by GSCMs with the column's parameters settings. Probabilities estimated from 100 generated contours. The GSCM parameter settings are referred to as  {\textit{Shape A}} (left), {\textit{Shape B}} (middle), and {\textit{Shape C}} (right). For all shapes $p = 50$ and $\kappa = 2.$ Values for $\boldsymbol{\mu}$ and $\boldsymbol{\sigma}$ for all shapes can be found in Supplement Section \ref{supp: simPars}.}
\label{fig: shapesAll}
\end{figure}

 Before defining the star-shaped representation, we review the standard definitions of a star-shaped polygon and its kernel \citep[][p. 18]{Preparata1985}. 
\begin{definition}{Star-shaped polygon:}
\label{def: starShapedPolygon}
A polygon $\boldsymbol{\underline{P}}$ is star-shaped if there exists a point $\boldsymbol{D}$ within $\boldsymbol{\underline{P}}$ such that the line segment $\overline{\boldsymbol{D}\boldsymbol{p}}$ is fully contained within $\boldsymbol{\underline{P}}$ for all points $\boldsymbol{p}$ on line $\boldsymbol{\overline{P}}$.
\end{definition}
All convex polygons are  star-shaped, but the set of star-shaped polygons is substantially larger. Figure \ref{fig: shapesAll} shows nine example star-shaped polygons.  

\begin{definition}{Kernel of a star-shaped polygon, $\mathcal{K}(\boldsymbol{\underline{P}})$:}
 The set of point(s) that satisfy the criterion for $\boldsymbol{D}$ in Definition \ref{def: starShapedPolygon} is referred to as the kernel of the polygon, $\mathcal{K}(\boldsymbol{\underline{P}})$.
\end{definition}
Convex polygons are the subset of star-shaped polygons such that $\mathcal{K}(\boldsymbol{\underline{P}}) = \boldsymbol{\underline{P}}$.

 For any star-shaped polygon, $\boldsymbol{\underline{S}}$, lines can be drawn from some point, $\boldsymbol{C}$, in the kernel of $\boldsymbol{\underline{S}}$ to all points on a contour, $\boldsymbol{\overline{S}}$. Assume for the moment that the contour point sequence $\boldsymbol{S}$ is unknown, but that the location of $\boldsymbol{C}$ is known along with the lengths and directions of the lines from $\boldsymbol{C}$ to $\boldsymbol{S}$. Then $\boldsymbol{S}$ could be derived from this information with trigonometry. Taking inspiration from this fact,  we develop the star-shaped representation.   First we define a line set:   
\begin{definition}{Line set, $\mathcal{L}(\boldsymbol{C}, \boldsymbol{\theta})$:}
A set of $p > 2$ lines, $\mathcal{L} = (\ell_{1}, \hdots, \ell_{p})$, extending infinitely outward from some starting point $\boldsymbol{C} = (C_{x}, C_{y})$ at $p$ unique angles, $\boldsymbol{\theta}  = (\theta_{1}, \hdots, \theta_{p})$, where $C_{x}$ and $C_{y}$ are x- and y-coordinates respectively. 
\end{definition}

\noindent We also define a set of spatial points that produce a star-shaped polygon when connected in order:

  \begin{definition}{Star-shaped point set, $\boldsymbol{V}( \boldsymbol{C}, \boldsymbol{\theta}, \boldsymbol{y})$:} 
  \label{def: starShapedRef}
  A set of $p > 2$ spatial points $\boldsymbol{V}= (\boldsymbol{v}_{1}, \hdots, \boldsymbol{v}_{p})$ such that 
\begin{equation}
\label{eq: calcSi}
\boldsymbol{v}_{i} = (C_{x} + y_{i}\cos(\theta_{i}), C_{y}  + y_{i} \sin (\theta_{i}))
\end{equation}
where $\boldsymbol{y} = (y_{1}, \hdots, y_{p})$ is a set of $p$ distances, $\boldsymbol{C} = (C_{x}, C_{y})$ is a spatial point, and $\boldsymbol{\theta} = (\theta_{1}, \hdots, \theta_{p})$ is a set of $p$ unique angles.
\end{definition}

\noindent  A star-shaped point set can be used to represent a contour when the distances are selected systematically:
\begin{definition}{Star-shaped representation, $\boldsymbol{\tilde{V}}(\boldsymbol{C}, \boldsymbol{\theta}, \boldsymbol{S}$)}: \label{def: starShapedApprox} Let  $\boldsymbol{\underline{S}}$ be a star-shaped polygon and  $\boldsymbol{C} \in \mathcal{K}(\boldsymbol{\underline{S}})$ be a starting point. Then, the star-shaped representation of  the contour $\boldsymbol{S}$, denoted by $\boldsymbol{\tilde{V}}(\boldsymbol{C}, \boldsymbol{\theta}, \boldsymbol{S})$,  is the star-shaped point set, $\boldsymbol{V}(\boldsymbol{C}, \boldsymbol{\theta}, \boldsymbol{y})$, where $\boldsymbol{y} = (y_{1}, \hdots, y_{p})$ is the set of distances from $\boldsymbol{C}$ to the intersection point of the contour line $\boldsymbol{\overline{S}}$ and each line $\ell_{i}$ in the line set  $\mathcal{L}( \boldsymbol{C}, \boldsymbol{\theta})$. 
  \end{definition}
  
The right panel of Figure \ref{fig: starVPoint} shows the components of the star-shaped representation for a sample contour.   Let $\boldsymbol{\overline{V}}$ refer to the contour line  formed by connecting $\boldsymbol{v}_{i}$ to $\boldsymbol{v}_{i + 1}$ for $i = 1, \hdots, p - 1$ and $\boldsymbol{v}_{p}$ to $\boldsymbol{v}_{1}$ for $p > 2$. Let $\boldsymbol{\underline{V}}$ refer to the polygon contained within $\boldsymbol{\overline{V}}$.

\begin{theorem}
\label{theorem: existsV}
Let $\boldsymbol{\theta} = (\theta_{1}, \hdots, \theta_{p})$ with $\theta_{i} < \theta_{i + 1}$ and $\theta_{i} \in (0, 2\pi)$ for all $i$. For a star-shaped polygon $\boldsymbol{\underline{S}}$ there exist $\boldsymbol{\theta}$ and $\boldsymbol{y}$ such that $\boldsymbol{\tilde{V}}(\boldsymbol{C}, \boldsymbol{\theta}, \boldsymbol{y}) = \boldsymbol{S}$ for any $\boldsymbol{C} \in \mathcal{K}(\boldsymbol{\underline{S}})$. (Proof in Appendix \ref{app: proofs}.)
\end{theorem}
\begin{corollary}
\label{corollary: distinctL}
Let $\ell_{\theta}$ denote the line that extends infinitely outward from $\boldsymbol{C}$ at angle $\theta \in (0, 2\pi)$ and that intersects $\boldsymbol{\overline{S}}$. For any $\theta$, the line $\ell_{\theta}$ is distinct, i.e. $\ell_{\theta} \neq \ell_{\theta'}$ for any $\theta, \theta'$ such that $\theta \neq \theta'$. (Proof in Appendix \ref{app: proofs}.)
\end{corollary}

The star-shaped representation allows for finding how contours differ and what the variability of the contours is in different spatial areas.  For example, consider two contours $\boldsymbol{S}_{k}$ and $\boldsymbol{S}_{\ell}$ described with star-shaped representations, $\boldsymbol{\tilde{V}}(\boldsymbol{C}, \boldsymbol{\theta}, \boldsymbol{S}_{k})$ and  $\boldsymbol{\tilde{V}}(\boldsymbol{C}, \boldsymbol{\theta}, \boldsymbol{S}_{\ell})$, for common line set $\mathcal{L}(\boldsymbol{C}, \boldsymbol{\theta})$. To find how much further one contour extends in any direction, simply find the difference between $y_{i, k}$ and $y_{i,\ell}$ where $y_{i, k}$ and $y_{i,\ell}$ are the distances from $\boldsymbol{C}$ to contours $\boldsymbol{S}_{k}$ and $\boldsymbol{S}_{\ell}$ along a line extending at angle $\theta_{i}$.   The variability of the contours along any line $\ell_{i} \in \mathcal{L}(\boldsymbol{C}, \boldsymbol{\theta})$ is estimated from the variability of the corresponding $\boldsymbol{y}_{i}$ values in the contours'  star-shaped representations.

For a contour enclosing a star-shaped polygon, the star-shaped representation is identical to the point-sequence representation when $p = n$, $\boldsymbol{C} \in \mathcal{K}(\boldsymbol{\underline{P}})$, and  $\theta_{i}$ aligns with the direction of the line segments $\overline{\boldsymbol{C}\boldsymbol{s}_{i}}$ for all $i$.  
When these conditions are met, the points $\boldsymbol{V}$ are the same for any choice of starting point $\boldsymbol{C}$ within the kernel of $\boldsymbol{\underline{S}}$. However, the angles $\boldsymbol{\theta}$ and lengths $\boldsymbol{y}$ will differ depending on $\boldsymbol{C}$.

\section{Star-shaped contour model}
\label{sec: GSCM}
\subsection{General model}
\label{sec: ssProbModel}
We now propose the {\it{star-shaped contour model}} for generating contours that enclose star-shaped polygons.\begin{definition}{Star-shaped contour model, $\boldsymbol{V}(\boldsymbol{C}, \boldsymbol{\theta}, \pi)$:} \label{def: starShapedProb} Let $\boldsymbol{C} = (C_{x}, C_{y})$ be a fixed starting point, let $\boldsymbol{\theta} = (\theta_{1}, \hdots, \theta_{p})$ be a fixed set of $p > 2$ unique angles, and let $\pi$ be a probability distribution from which a set of values $\boldsymbol{y}  = (y_{1}, \hdots, y_{p})> 0$ can be drawn.  These parameters form a  star-shaped probability model if drawn sets $\boldsymbol{y}$ are used to form corresponding star-shaped points sets,  $\boldsymbol{V}(\boldsymbol{C}, \boldsymbol{\theta}, \boldsymbol{y})$, as given in Definition \ref{def: starShapedRef}. 
\end{definition}

We now consider a distribution, $\pi$, that is appropriate in many circumstances. We assume that $\boldsymbol{y}$ follows a Gaussian distribution,
\begin{equation}
\label{eq: normDist}
\boldsymbol{y} \sim N(\boldsymbol{\mu}, \bold{\Sigma}) ,
\end{equation}
where $\boldsymbol{\mu}$ is a mean vector and the parameter $\bold{\Sigma}$ is a positive-definite covariance function.  We further assume that $\boldsymbol{\mu}$ and $\bold{\Sigma}$ are such that mass on non-positive $\boldsymbol{y}$ is negligible.   (In practice, if in a small proportion of cases, a generated $y_{i}$ is non-positive,  its value can be set to some small $\eta >0$.) We call this model the {\it Gaussian Star-shaped Contour Model} (GSCM). The  GSCM can be seen as a finite approximation to the planar version of Gaussian Random Particles proposed in \citet{Hansen2015}. GSCMs can produce a fairly flexible set of contours. The first three rows in Figure \ref{fig: shapesAll} illustrates the types of contours GSCMs can produce.

 Because of how $\boldsymbol{y}$ is constructed, reasonable $\boldsymbol{\mu}$  and $\bold{\Sigma}$ that align with typical observable $\boldsymbol{y}$ values will avoid substantial non-positive $\boldsymbol{y}$. The values $\boldsymbol{y}$ represent distances from a starting point $\boldsymbol{C}$, so are automatically non-negative. Some points in the kernel of the polygon will typically be centrally located points. Lengths close to zero can be avoided by using one of these points  for $\boldsymbol{C}$. 
 
Covariance matrices, $\bold{\Sigma}$, are based on the structure of the lines in the line set.  The correlation structure for the set of distances, $\boldsymbol{y}$, is a function of the angles, $\boldsymbol{\theta}$, of the lines in the line set. Covariances based on angles are complicated by the fact that  $0$ and $2\pi$ represent the same angle. So, the difference between two angles does not necessarily correspond to how far apart the angles actually  are. Specialized covariance functions have been derived that remain  valid when distances are indexed by angle \citep{Gneiting2013}. 
Denote the angle between $\theta_{i}$ and $\theta_{j}$ by
$d(\theta_{i}, \theta_{j}) \in [0, \pi]$.

Typically, the correlation between $y_{i}$ and $y_{j}$ will  decrease as $d(\theta_{i}, \theta_{j})$ increases. 
For the simulation examples in this paper,  we focus on an exponential covariance structure, $\bold{\Sigma}(\boldsymbol{\sigma}, \kappa)$ where $\boldsymbol{\sigma} = (\sigma_{1}, \hdots, \sigma_{p})$ and $\kappa > 0$. The element, $\bold{\Sigma}_{ij}$, in the $i$-th row and $j$-th column of this covariance is
\begin{equation}
\label{eq: expCov}
\bold{\Sigma}_{ij} = \sigma_{i} \sigma_{j} \exp \left(- \frac{d(\theta_{i}, \theta_{j})}{\kappa}   \right). 
\end{equation}

\subsection{Fitting GSCMs}
\label{sec: fitting}

We now turn  to building a GSCM given observed contours. We assume that the data are $N$ observed contours, $\bold{S} = (\boldsymbol{S}_{1}, \hdots, \boldsymbol{S}_{N})$,  that enclose regions that are star-shaped polygons. We also assume that the contours are generated from a common, but unknown, $\boldsymbol{C}$ and $\boldsymbol{\theta}$, i.e.,  $\boldsymbol{C} = \boldsymbol{C}_{1} = \cdots \boldsymbol{C}_{N}$ and $\boldsymbol{\theta} = \boldsymbol{\theta}_{1}, \cdots \boldsymbol{\theta}_{N}$. 
 We first find a starting point, $\boldsymbol{\hat{C}}$, and angles, $\boldsymbol{\hat{\theta}}$. Then, we estimate $\boldsymbol{\mu}$ and the parameters controlling $\bold{\Sigma}(\cdot)$ based on the observed $\boldsymbol{y}$ for the selected $\boldsymbol{\hat{C}}$ and angles $\boldsymbol{\hat{\theta}}$.

\subsubsection{Fixing the starting point {$\hat{\boldsymbol{C}}$} and the set of angles {$\hat{\boldsymbol{\theta}}$} }
\label{sec: determineCAndP}
We fix a starting point,  $\boldsymbol{\hat{C}}$, and set of angles, $\boldsymbol{\hat{\theta}}$,  that can be used to quantitatively describe the observed contours accurately. We first describe how to find $\hat{\boldsymbol{C}}$ conditional on $\boldsymbol{\theta}$ and $\boldsymbol{\hat{\theta}}$ conditional on $\boldsymbol{C}$ separately. Then, we describe an iterative algorithm to fix both values together.\\

\noindent \textit{Finding $\hat{\boldsymbol{C}}$ conditional on $\boldsymbol{\theta}$: } The fixed starting point, $\boldsymbol{\hat{C}}$, is selected to minimize the difference in area between the observed contours' enclosed polygons and the star-shaped representations of the observed contours' enclosed polygons.  Conditional on $\boldsymbol{\theta}$, we define the set that differs between an observed polygon, $\boldsymbol{\underline{S}}_{i}$, and a star-shaped representation of that contour with  starting point, $\boldsymbol{\hat{C}}$, as
 \begin{align}
A (\boldsymbol{\hat{C}}, \boldsymbol{\theta}, \boldsymbol{S}_{i}) &:= \{ (\underline{\boldsymbol{S}_{i}}^{c} \cap \underline{\boldsymbol{\tilde{V}}}(\boldsymbol{\hat{C}}, \boldsymbol{\theta}, \boldsymbol{S}_{i})  \cup (\underline{\boldsymbol{S}}_{i} \cap  \underline{\boldsymbol{\tilde{V}}}^{c}(\boldsymbol{\hat{C}}, \boldsymbol{\theta}, \boldsymbol{S}_{i}))\}.
\end{align}
where the superscript $c$ denotes the complement of the set. The area contained within this set is denoted by $|A (\boldsymbol{\hat{C}}, \boldsymbol{\theta}, \boldsymbol{S}_{i})|$.  
Assuming $\boldsymbol{\theta}$ is known,  Theorem \ref{theorem: existsV} guarantees that at least one point $\boldsymbol{\hat{C}}$ exists where $|A(\boldsymbol{C}, \boldsymbol{\theta}, \boldsymbol{S}_{i}) | = 0$ for all $i$. So, the best selection of $\boldsymbol{\hat{C}}$ would be the point that gives  $|A (\boldsymbol{\hat{C}}, \boldsymbol{\theta}, \boldsymbol{S}_{i})| = 0$ for all $i$, i.e., $\boldsymbol{\hat{C}} = \boldsymbol{C}$. However, finding the difference in area  at all possible locations is not compuationlly feasible, so $\boldsymbol{\hat{C}}$ is  numerically approximated as
\begin{align}
\label{eq: CHatOptD}
\boldsymbol{\hat{C}} = \underset{\bold{C} \in \boldsymbol{\mathcal{D}} }{\text{argmin}} \left \{ f ( |A (\boldsymbol{C}, \boldsymbol{\theta}, \boldsymbol{S}_{1})|, \hdots, |A (\boldsymbol{C}, \boldsymbol{\theta}, \boldsymbol{S}_{N})|)\right\} , 
\end{align}
where $f$ is the mean function and $\boldsymbol{\mathcal{D}}$ is a grid of points. The finer the grid of $\boldsymbol{\mathcal{D}}$ the closer $\boldsymbol{\hat{C}}$ will be to $\boldsymbol{C}$. The number of points in $\boldsymbol{\mathcal{D}}$ can be reduced by considering the kernels of star-shaped polygons.  
\begin{theorem}
\label{theorem: startPoint}
 Let  $\bold{\underline{S}}$ be a set of $N$ star-shaped polygons, let $\boldsymbol{C}$ be the true starting point, and let $\boldsymbol{\mathcal{\hat{K}}}(\bold{\underline{S}}) = \mathcal{K}(\boldsymbol{\underline{S}}_{1}) \cap \hdots \cap \mathcal{K}(\boldsymbol{\underline{S}}_{N})$ denote the intersection of the kernels of all polygons. Then, $\boldsymbol{C} \in \boldsymbol{\hat{\mathcal{K}}}(\bold{\underline{S}})$. (Proof in Appendix \ref{app: proofs}.)
\end{theorem}

\noindent Therefore, we need only perform the optimization in Equation \ref{eq: CHatOptD} for  $\boldsymbol{\mathcal{D}} \in \boldsymbol{\hat{\mathcal{K}}}(\bold{\underline{S}})$. Algorithms for computing $\boldsymbol{\hat{\mathcal{K}}}(\bold{\underline{S}})$ are given in Appendix \ref{app: compK}.\\

\begin{figure}[h!]
\centering
\includegraphics[width=.7\textwidth]{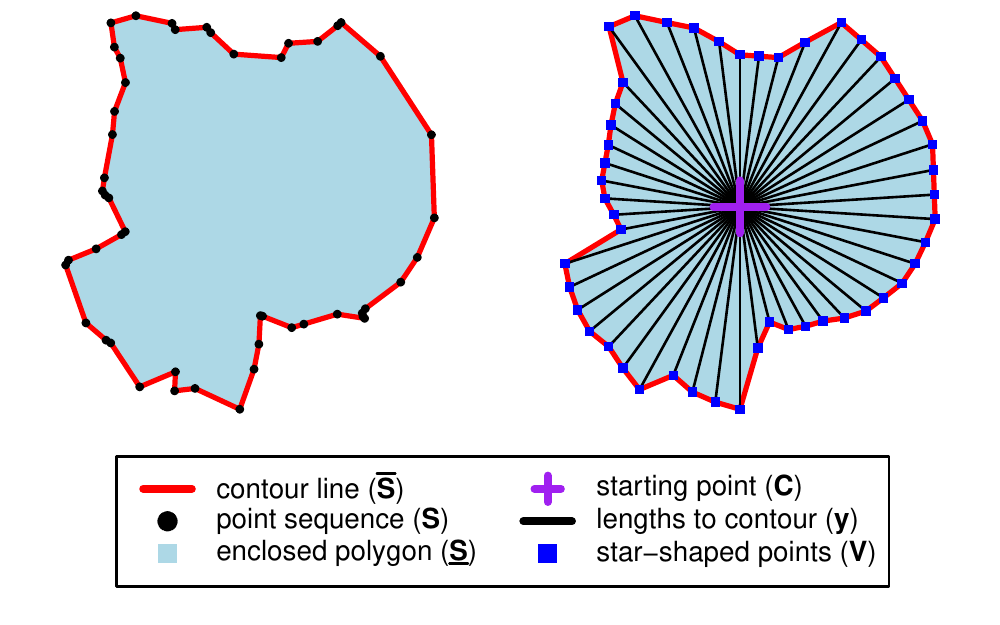}
\caption{Components of a contour represented by a point-sequence representation (left) and a star-shaped representation approximating the point sequence with evenly-spaced angles (right).} 
\label{fig: approxContour}
\end{figure}

\noindent \textit{Finding $\boldsymbol{\hat{\theta}}$ conditional on $\boldsymbol{C}$: }
\label{sec: detPHat}
The set of angles, $\boldsymbol{\hat{\theta}}$, is selected to keep the mean difference in area between the observed contours' enclosed polygons and  the star-shaped representations  of these contours' enclosed polygons below some value.  Following Corollary \ref{corollary: distinctL}  any set of distinct angles can be used to form a star-shaped representation of a contour. For simplicity, we recommend using evenly spaced angles. Setting $\boldsymbol{\hat{\theta}}$  then reduces to finding $\hat{p}$, the number of elements in $\boldsymbol{\hat{\theta}}$.
Figure \ref{fig: approxContour} illustrates how a star-shaped contour is approximated with a star-shaped representation with evenly-spaced $\boldsymbol{\theta}$.
 
Since $\hat{p}$ controls the dimensions of $\boldsymbol{\mu}$ and $\bold{\Sigma}$, larger $\hat{p}$ requires more computation.  We then identify the approximately lowest $\hat{p}$ that keeps the mean difference in area below some value. To allow comparisons of polygons of different sizes, we often express this allowable mean difference in area as a proportion, $\delta$, of the average area of the polygons. This constraint on the allowable differing area is then 
\begin{align}
\label{eq: redP}
f (|A (\boldsymbol{C}, \boldsymbol{\hat{\theta}}, \boldsymbol{S}_{1})|, \hdots, |A (\boldsymbol{C}, \boldsymbol{\hat{\theta}}, \boldsymbol{S}_{N})| ) < \frac{\delta}{N} \sum_{i=1}^{N} |\boldsymbol{\underline{S}_{i}}| ,
\end{align}
where $f$ is the mean function and $|\boldsymbol{\underline{S}_{i}}|$ denotes the area of the polygon $\boldsymbol{\underline{S}}_{i}$.   We use Algorithm \ref{algo: findTheta} to find an approximately minimal $\hat{p}$ that satisfies the constraint in Equation \ref{eq: redP}.  To avoid selecting a larger $\hat{p}$ than necessary, $p^{(0)}$ should generally be initialized such that Equation \ref{eq: redP} is not  satisfied. Smaller values of $a$ will make $\hat{p}$ more precise, but will require more computation than larger $a$. Using lower values of $\delta$ will generally result in higher $\hat{p}$ and lower differences in area. How to balance computation time and the appropriate value of $\delta$ is informed by the application of interest.  Section \ref{sec: varyingDelta} uses simulation to explore different $\delta$. \\

\begin{algorithm}
\caption{Finding $\boldsymbol{\hat{\theta}}$ conditional on $\boldsymbol{C}$}
	Initialize $p^{(0)}$ and set $t \leftarrow 0$ \\
	Compute RHS of Equation \ref{eq: redP}\\	
    \While{Equation \ref{eq: redP} does not hold}{
    Compute $\boldsymbol{\theta}^{(t)}$ for $p^{(t)}$\\
    Compute LHS of Equation \ref{eq: redP} with $\boldsymbol{\theta} = \boldsymbol{\theta}^{(t)}$ \\
   	 \eIf{Equation \ref{eq: redP} does not hold}{
	 	Set $t \leftarrow t + 1$ \\
	   Set $p^{(t+1)} \leftarrow ap^{(t)}$ where is $a > 1$}
	   {Set $\boldsymbol{\hat{\theta}} \leftarrow \boldsymbol{\hat{\theta}}^{(t)}$}
 }
 \label{algo: findTheta}
 \end{algorithm}

\newpage
\noindent \textit{Finding $\boldsymbol{\hat{C}}$ and $\boldsymbol{\hat{\theta}}$:} In practice, neither $\boldsymbol{\hat{C}}$ nor $\boldsymbol{\hat{\theta}}$ will be known. So, to find both values, we iterate between setting $\boldsymbol{\hat{C}}$ conditional on $\boldsymbol{\hat{\theta}}$ and $\boldsymbol{\hat{\theta}}$  conditional on $\boldsymbol{\hat{C}}$. Algorithm \ref{algo: findThetaP} describes this process. As in Algorithm \ref{algo: findTheta}, the initial value, $p^{(0)}$, should be selected to be low enough that we are confident that Equation \ref{eq: redP} is not satisfied. Otherwise, we may select a larger $\hat{p}$ than is needed.  Smaller values of $a$ will result in more precise determination of $\hat{p}$, but will require more computation.  \\

\begin{algorithm}
\caption{Finding $\boldsymbol{\hat{C}}$ and $\boldsymbol{\hat{\theta}}$}
	Initialize $p^{(0)}$ and set $t \leftarrow 0$ \\
	Compute RHS of Equation \ref{eq: redP}\\	
    \While{Equation \ref{eq: redP} does not hold}{
    Compute $\boldsymbol{\theta}^{(t)}$ for $p^{(t)}$\\
    Find $\boldsymbol{\hat{C}}^{(t)}$ using Equation \ref{eq: CHatOptD} with $\boldsymbol{\theta} = \boldsymbol{\theta}^{(t)}$\\
    Compute LHS of Equation \ref{eq: redP} with $\boldsymbol{C} = \boldsymbol{\hat{C}}^{(t)}$ and $\boldsymbol{\theta} = \boldsymbol{\hat{\theta}}^{(t)}.$ \\
   	 \eIf{Equation \ref{eq: redP} does not hold}{
	 	Set $t \leftarrow t + 1$ \\
	   Set $p^{(t+1)} \leftarrow ap^{(t)}$ where is $a > 1$}
	   {Set $\boldsymbol{\hat{C}} \leftarrow \boldsymbol{\hat{C}}^{(t)}$\\
	   Set $\boldsymbol{\hat{\theta}} \leftarrow \boldsymbol{\hat{\theta}}^{(t)}$}
 }
 \label{algo: findThetaP}
\end{algorithm}

\subsubsection{Computing a posterior}
\label{sec: compPost}
Once $\boldsymbol{\hat{C}}$ and $\boldsymbol{\hat{\theta}}$ determined, model fitting is straightforward. For each observed contour, the observed $\boldsymbol{y}$ values are computed given $\boldsymbol{\hat{C}}$ and $\boldsymbol{\hat{\theta}}$.  Then from Equation \ref{eq: normDist}, the corresponding likelihood for these $\bold{y}$ is just that of a multivariate normal distribution:
\begin{align}
\prod_{j=1}^{N} (2\pi)^{-\hat{p}/2} \det \left(\bold{\Sigma}(\cdot) \right)^{-1/2} \exp \left\{ - \frac{1}{2} (\boldsymbol{y}_{j} - \boldsymbol{\mu})^{T} \bold{\Sigma}(\cdot)^{-1}(\boldsymbol{y}_{j} - \boldsymbol{\mu}) \right\} .
\end{align}
Estimates of the mean vector $\boldsymbol{\mu}$ and the parameters controlling $\bold{\Sigma}(\cdot)$ can be estimated using any standard method.

For demonstration purposes in the simulations and example that follow we take a Bayesian approach. We assume an exponential covariance as in Equation \ref{eq: expCov} with parameters $\boldsymbol{\sigma}$ and $\kappa$. We use the following simple prior distributions for $\boldsymbol{\mu}$, $\boldsymbol{\sigma}$ and $\kappa$. We assume a multivariate normal prior distribution for $\boldsymbol{\mu}$:
\begin{align}
\boldsymbol{\mu} & \sim \text{MVN}(\boldsymbol{\mu_{0}}, \bold{\Lambda}_{0}).
\end{align}
The hyperparameters  $\mu_{0}$ and $\bold{\Lambda}_{0}$  give the mean and covariance of the prior distribution. For the example exponential covariance defined in Equation \ref{eq: expCov} we assume uniform priors on $\kappa$ and $\boldsymbol{\sigma}$:
 \begin{align}
 \kappa  & \sim  \text{Unif}(0, \beta_{\kappa, 0})\\
  \sigma_{j} & \sim \text{Unif}(0, \beta_{\sigma, 0})
\end{align}
where the hyperparameters $\beta_{\kappa, 0}$ and $\beta_{\sigma, 0}$ are upper bounds. Samples from the posterior distributions of the parameter can be found via standard Markov chain Monte Carlo (MCMC).

\subsubsection{Estimating gridded probabilities and credible intervals}
\label{sec: probsCredInts}
Sampled contours can be used to estimate the probability of a given area being contained within a contour. Consider $K$ sample contours, $\bold{\hat{S}} = (\boldsymbol{\hat{S}}_{1}, \hdots \boldsymbol{\hat{S}}_{K})$.  Each generated contour can be approximated by a binary grid, $\bold{G}$, of dimension $r \times v$. Let $g_{i, j}$ indicate the grid box in the $i$-th row and $j$-th column of $\bold{G}$. Let $\mathbbm{1}{g_{i,j,k}}$ indicate whether the majority of the area in grid box $g_{i,j}$ is inside or outside contour $k$. Most grid boxes will be entirely inside or outside the contour; however, grid boxes that intersect the generated contour will contain area both inside and outside the contour. Ideally, the grid selected should be fine enough to ensure that little area is contained within these transitional grid boxes.  Averaging the binary grids produces an $r \times v$ matrix, $\bold{\hat{P}}$,  with elements $\hat{p}_{i,j} = \sum_{i=1}^{K}\mathbbm{1}{g_{i,j,k}} / K$, that indicate  the probability of grid box $g_{i,j}$ being contained within a contour. The last row of Figure  \ref{fig: shapesAll} shows estimated gridded probabilities obtained from $K = 100$ generated contours from the corresponding GSCMs. 

 Credible regions for the location of the contour can be computed from $\bold{\hat{P}}$. The $(1 - \alpha)$ credible region, $\boldsymbol{I}_{1 - \alpha}$, is formed from a union of grid boxes that satisfy the condition
\begin{equation}
\label{eq: credReg}
\boldsymbol{I}_{1 - \alpha} = \left\{ g_{i,j}: \frac{\alpha}{2} < \hat{p}_{i, j} < 1 - \frac{\alpha}{2} \right\}.
\end{equation}

\subsubsection{Rescaling data}
\label{sec: rescaleData}
For numerical convenience in fitting and generating contours, it is often desirable for all contours to be contained within the $[0, 1] \times [0, 1]$ unit square. Observed data will typically need to be rescaled to be within these bounds. Data should be rescaled such that generated contours do not extend outside the unit square. A good rescaling also ensure that the contours that will be generated rarely, if ever,  extend outside the unit square. 

Therefore, we rescale observed contours $\bold{S} = (\boldsymbol{S}_{1}, \hdots \boldsymbol{S}_{N})$  to be within  an  $[\epsilon, 1 - \epsilon]  \times [\epsilon, 1 - \epsilon]$ square. This rescaling provides a buffer region of width $\epsilon$ on the outside of the unit square in which no contours have been observed.  Therefore, if generated contours extend farther than the observed contours, they will typically go into this buffer region rather than outside the unit square.  The higher the variability of contours, the larger the value of $\epsilon$ needed to avoid generating contours that go beyond the unit square. 

To translate a set of observed coordinates, $\bold{S} = (\boldsymbol{S}_{1}, \hdots, \boldsymbol{S}_{n})$,  to the  square of dimension $[\epsilon, 1 - \epsilon]  \times [\epsilon, 1 - \epsilon]$, let $\min(\bold{S}_{x})$ and $\max(\bold{S}_{x})$ denote the minimum and maximum observed $x$-coordinates from all spatial points in all contours in $\bold{S}$. Define $\min(\bold{S}_{y})$ and  $\max(\bold{S}_{x})$ analogously for the $y$-coordinates. Let $\boldsymbol{s}_{i,j} = (s_{i,j}^{x}, s_{i,j}^{y})$ denote the $i$-th point  in the $j$-th  contour. Then for all $i$ and all $j$, the equivalent rescaled coordinates for observed points $\boldsymbol{s}_{i,j}$ are 
\begin{align}
\tilde{s}_{i,j}^{x} = \epsilon + (1 - 2\epsilon) (s_{i,j}^{x} - \min(\bold{S}_{x}))/( \max(\bold{S}_{x}) - \min(\boldsymbol{S}_{x})) , \\
\tilde{s}_{i,j}^{y} = \epsilon + (1 - 2\epsilon) (s_{i,j}^{y} - \min(\bold{S}_{y}))/( \max(\bold{S}_{y}) - \min(\bold{S}_{y})).
\end{align}

\section{Coverage Metric}
\label{sec: coverageMetric}
To assess if a probabilistic contour model performs well, a metric is needed. A good model correctly identifies the region where the contour could plausibly be located. So, we focus on the coverage of prediction intervals for star-shaped contours.  
With an accurate contour model, the variability of the generated contour would be correctly represented along all parts of the contour.  In designing an appropriate metric, we leverage the  star-shaped structure of the data. The general idea is to assess coverage for  each line in a line set individually.

\begin{figure}[h!]
\centering
\includegraphics[width=.75\textwidth]{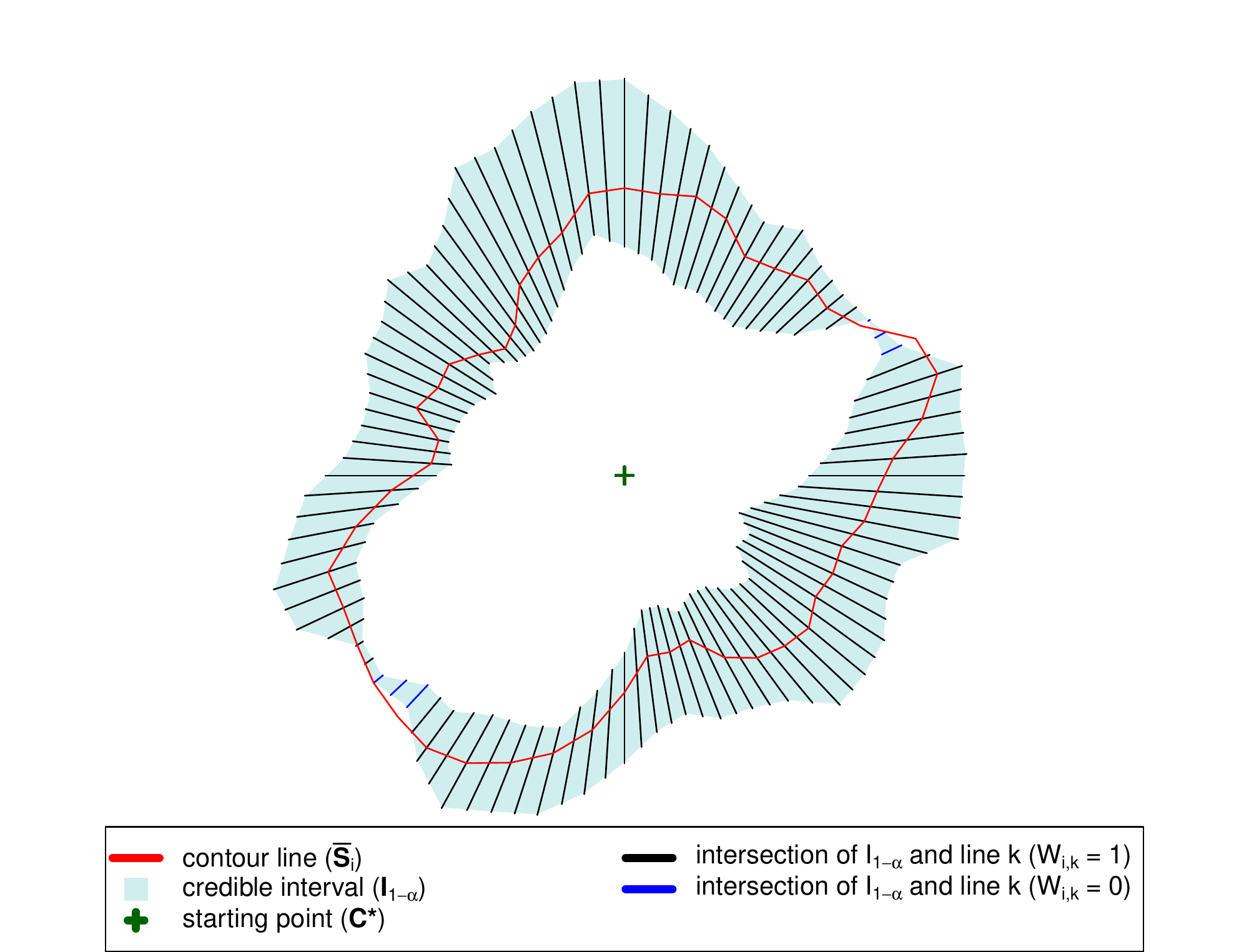}
\caption{Illustration of coverage assessment for a contour line, $\boldsymbol{\overline{S}}_{i}$ (red),  and a $1-\alpha$ credible region, $\boldsymbol{I}_{1 - \alpha}$ (light blue). The line segments, $I_{1 - \alpha, k}$,  corresponding to the intersection of the $\boldsymbol{I}_{1 - \alpha}$ credible region and line $\ell_{k}$  are colored black when they cover $S_{i, k}$ and blue otherwise. The center of the star-shaped polygon from which the contour is generated is denoted by a green cross sign.   } 
\label{fig: metricDemo}
\end{figure}

To make this idea precise, we define several quantities illustrated in Figure \ref{fig: metricDemo}. As in Equation \ref{eq: credReg}, let $\boldsymbol{I}_{1 - \alpha}$ be the $1-\alpha$ credible region obtained from some contour model. Define some test line set $\mathcal{L}^{*}(\boldsymbol{C}^{*}, \boldsymbol{\theta}^{*})$ with $M$ evenly-spaced lines. Define $I_{1-\alpha, k}$ as the line segment formed from the intersection of the $\boldsymbol{I}_{1 - \alpha}$ credible region and the line $\ell_{k} \in \mathcal{L}^{*}(\boldsymbol{C}^{*}, \boldsymbol{\theta}^{*})$.  We refer to $I_{1-\alpha, k}$  as a test line. Also, define $R_{i, k}$ as the intersection of some observed contour $\boldsymbol{S}_{i}$ and line $\ell_{k}$. 
Note that $R_{i,k}$ will always be a single point when the polygon $\boldsymbol{\underline{S}}$ is exactly star-shaped and $\boldsymbol{C}^{*} \in \mathcal{K}(\underline{\boldsymbol{S}_{i}})$.  

Let $W_{i, k} = \mathbbm{1}[R_{i, k} \in I_{1 - \alpha, k}]$  indicate whether the intersection points of the observed contour and the line $\ell_{k} \in \mathcal{L}^{*}(\boldsymbol{C}^{*}, \boldsymbol{\theta}^{*})$ are contained within the intersection of the credible region and line $\ell_{k}$.  Then, for credible intervals with perfect coverage, for any $i, k$,
\begin{equation}
Pr(W_{i, k}) =  \mathbb{E}[W_{i, k}] = 1 - \alpha.
\end{equation}
  In other words, the part of the true contour that intersects line $\ell_{k}$ is contained within the credible contour region along line $\ell_{k}$ with probability $1- \alpha$.

We consider coverage behavior for a set of $N$ observed contours, $\bold{S} = (\boldsymbol{S}_{1}, \hdots, \boldsymbol{S}_{N})$ that enclose star-shaped polygons. Each $\boldsymbol{S}_{i}$ is assumed to be independent of $\boldsymbol{S}_{j}$ for all $i, j$.

For a credible interval with perfect coverage  for any $i$, $k$,
 \begin{equation}
 \label{eq: sampleCoverage}
  \sum_{i = 1}^{N} W_{i, k} \approx N(1 - \alpha).
  \end{equation}
for sufficiently large sample size, $N$. This equation is used to assess coverage in practice. Equation \ref{eq: sampleCoverage}  not  holding for any $k$ indicates the variability of the contour on test line $\ell_{k}$ is not correctly represented by the credible region.

With this set-up, $W_{i, k}$ and $W_{j, k}$ are independent for all $i, j$. So, for a given line $\ell_{k}$, 
\begin{equation}
\sum_{i=1}^{N}W_{i, k} \sim \text{Binomial} (N, 1 - \alpha).
\end{equation}
Since the distributional behavior of the quantity $\sum_{i=1}^{N}W_{i, k}$ is known, the expected variability of the mean coverage given a particular sample size is known. This information can be accounted for in a simulation study or cross-validation experiment.

We also consider how coverage on one test line in $\mathcal{L}^{*}(\boldsymbol{C}^{*}, \boldsymbol{\theta}^{*})$ relates to coverage on another test line. The location of points on a contour are generally correlated. So, $W_{i, k}$ and $W_{i, \ell}$ are correlated. Typically $W_{i, k}$ and $W_{i, \ell}$ will be more correlated the smaller the angle distance from $\theta_{k}$ and $\theta_{\ell}$. What this means is that while  
\begin{equation}
\mathbb{E} \left[ \sum_{k=1}^{M} W_{i, k} \right] =  M(1 - \alpha),
\end{equation}
is fixed and known, the quantity $\sum_{k=1}^{M} W_{i, k}$ is not a good metric to assess to coverage. The distribution of $\sum_{k=1}^{M} W_{i, k}$ depends on the correlation structure of the points on the contour. For a contour with high correlation among points, the quantity  $\sum_{k=1}^{M} W_{i, k}$  will be  substantially affected by what contour happens to be sampled. For intuition, consider a contour with high correlation among all $i,j$. In this case, the contour is likely to be entirely within the credible interval or entirely outside of it. So, $ \sum_{k=1}^{M} W_{i, k}$ will be either 0 or M. 

These relationships among coverage for $M$ and $N$ show that sample sizes need to be considered  in terms of the number of contours observed. The metric $\sum_{i=1}^{N} W_{i, k}$ should be considered for all elements in $\boldsymbol{\theta}$.   The number of test lines, $M$, should be set  such that accuracy is assessed with detail appropriate for the application. Since the true exact value of $p$,  is unknown, we cannot simply set $M  = p$.  However, based on the observed data, we should have a general idea of $p$. So, we set $M >> p$, to ensure that $M > p$. We also evenly space $\boldsymbol{\theta}^{*}$ with $\theta^{*}_{1} = (2\pi/M)/2 = \pi/M$.   By assessing coverage on a substantially greater number of test lines than the true number of lines, we ensure that coverage is at least assessed near every true line.  

To carry out this assessment, a fixed starting point $\boldsymbol{C}^{*}$ and $\boldsymbol{\theta}^{*}$ should be selected. In simulation studies, the true value of $\boldsymbol{C}$ will be known and we can let $\boldsymbol{C}^{*} = \boldsymbol{C}$. For assessment of real data such as in a cross-validation study, a starting point for $\boldsymbol{C}^{*}$ will be unknown and must be determined. We recommend using a $\boldsymbol{C}^{*}$  that minimizes the difference in area between the observed contours' enclosed polygons and the star-shaped representations of the observed contours'  enclosed polygons as in Section \ref{sec: determineCAndP}, i.e.,  let $\boldsymbol{C}^{*} = \boldsymbol{\hat{C}}$.
 
This assessment approach differs from how contours  have been assessed in the context of level exceedances. There, a credible region or confidence region has often been defined as the region that covers the true contour in its entirety $(1-\alpha)$-proportion of the time \citep[e.g.][]{Bolin2015, French2014}. With credible (confidence) regions constructed to satisfy this definition, coverage can be assessed by determining what proportion of the time the true contour is fully contained within the region. We opt not to use this metric since our goal is to develop a method to generate contours directly. Correlation along the contour makes assessing the probability of capturing the entire contour difficult. Our metric reflects that we are most concerned with getting the right variability in all parts of the contour. We are less concerned with identifying a larger area that contains that entirety of the contour with high probability.  Our intervals are therefore narrower than would be required for these global intervals.

\section{Simulation studies}
\label{sec: sims}

\subsection{Simulation details}
\label{sec: simDetails}
In the following simulation studies, we consider how the star-shaped model performs in inferring distributions of contours from datasets of observed contours. We consider performance with varying numbers of observations, different constraints for the allowable mean difference in area ($\delta$ as defined in Section \ref{sec: detPHat}), and varied GSCM parameters. 

In many of our simulations, we focus on a particular GSCM with $p = 50$  that we will refer to as {\textit{Shape A}}.  The correlation structure of $\boldsymbol{y}$ follows the exponential form given in Equation \ref{eq: expCov}. The vector of mean distances, $\boldsymbol{\mu}$, and variance parameter vector, $\boldsymbol{\sigma}$, change gradually.   The exact values of  $\boldsymbol{\mu}$ and $\boldsymbol{\sigma}$  can be found in Supplement Section \ref{supp: simPars}. Unless otherwise noted, $\kappa$ is set to  $2$.  Example generated contours and gridded probability estimates for {\it{Shape A}} are given in the left panel of Figure \ref{fig: shapesAll}.  

The values $\boldsymbol{\hat{C}}$ and $\boldsymbol{\hat{\theta}}$ are found as described  in Section \ref{sec: fitting}. The parameter values are fit  from observations using MCMC. Chains are run for 50,000 iterations with the first 15,000 iterations discarded as burn-in. The prior parameters are $\ \boldsymbol{\mu_{0}} = (0.2, \hdots, 0.2)$, $\beta_{\kappa_{0}} = (8, \hdots, 8)$, $\beta_{\sigma_{0}} = (0.15, \hdots, 0.15)$, and $\bold{\Lambda}_{0} = .05 \bold{I}_{\hat{p}}$ where $\bold{I}_{\hat{p}}$ is a diagonal matrix of dimension $\hat{p}$ by $\hat{p}$.  The value $\hat{p}$ refers to the number of angles in $\boldsymbol{\hat{\theta}}$. 

For all simulations, we use 40 evaluation runs. On each evaluation run, we estimate the GSCM parameters using $N$ contours generated from the true GSCM as training data. From the resulting fitted GSCM, we generate 100 contours and find credible intervals as described in Section \ref{sec: probsCredInts}. To evaluate coverage, we compare the credible interval with a single ``true" contour drawn from the true GSCM. We record the coverage for a set of $M = 100$ evenly spaced test lines with $\theta_{1}^{*} = \pi/M$ as described in Section \ref{sec: coverageMetric}. We report the mean coverage over the 40 evaluations and the standard deviation  across the $M = 100$ test lines.

\subsection{Varying number of observations, $N$}
\label{sec: nObs}
Our first simulation varies $N$, the number of simulated ``observed" contours used to fit the \textit{Shape A} GSCM.  We consider coverage performance for 10, 20, and 50 simulated observed contours with $\delta = 0.02$.    Table  \ref{tab: nObs} displays the results of this simulation.

We find that coverage improves for $N = 20$ compared to $N = 10$. We find slightly worse performance for $N = 50$ than $N= 20$, although the performances are not substantially different given that we had only 40 evaluation runs. These results indicate that obtaining some minimum sample size is important for coverage performance. For small sample sizes, on the order of $N = 10$, the data alone may not be enough to produce accurate coverage, particularly for the 80$\%$ credible interval. However, data sets of this size can potentially still be modeled correctly if informative priors can supplement the observations.

\begin{table}
\center
\caption{Mean coverage values for 40 simulations of fitting the contour distribution for \textit{Shape A} with different number of observed contours sampled as training data. In each simulation,  $M = 100$ evenly-spaced test lines were evaluated with $\theta^{*}_{1} = \pi/M$.  Standard deviations across the test lines are given in parentheses. Priors and MCMC details are given in Section \ref{sec: simDetails} }
\begin{tabular}{r|rrr}
  \hline
   \textbf{Nominal} & \textbf{N = 10} & \textbf{N =  20} & \textbf{N = 50} \\ 
  \hline
0.80 & 0.87 (0.04) & 0.79 (0.07) & 0.76 (0.06) \\ 
0.90 & 0.94 (0.04) & 0.89 (0.05) & 0.86 (0.05) \\ 
 0.95 & 0.98 (0.02) & 0.95 (0.03) & 0.93 (0.04) \\ 
   \hline
\end{tabular}
\label{tab: nObs}
\end{table}

\subsection{Varying allowable difference in area, $\delta$}
\label{sec: varyingDelta}
In these simulations, we evaluate how coverage accuracy is affected by the parameter $\delta$ introduced in  Section \ref{sec: detPHat}. This parameter controls the allowable differing area when setting the number of lines used in fitting, $\hat{p}$, and how accurate $\boldsymbol{\hat{C}}$ must be.  We evaluate coverage for  \textit{Shape A} with $\delta$ set to $0.03, 0.02$, and $0.01$. These $\delta$ selections set the allowable mean difference in area to 3$\%$, $2\%$, and $1\%$ of the mean area contained within the observed contours. We also consider how  correlation in $\boldsymbol{y}$ affects the need for different $p$ by evaluating each $\delta$ for three different $\kappa$ values: 1, 2, and 4. Table \ref{tab: variedDelta} displays the mean coverage across test lines for three $\alpha$-levels along with the mean $\hat{p}$ found. On each evaluation the number of sampled contours is set to $N = 20$.

We find that the mean coverage accuracy is only modestly affected by the value of $\delta$. These results support the idea that using a lower $\delta$ in many cases will reduce computation while not reducing model performance. We also find that, for a given $\delta$, an increase in $\kappa$ corresponds to a decrease in $\hat{p}$. In other words, for a contour with higher correlation among $\boldsymbol{y}$, a smaller set of lines can adequately represent  the contour distribution.

\begin{table}
\caption{Mean coverage values for 40 simulations fitting the contour distribution for \textit{Shape A} with different values of $\delta$. In each simulation,  20 observed contours were sampled  as training data and $M = 100$ evenly-spaced test lines with $\theta^{*} = \pi/M$ were evaluated.  Standard deviations across the test lines are given in parentheses. The mean $\hat{p}$ is given for each $\delta$ along with the standard deviation across the evaluation runs in parentheses. Apostrophes indicate that the entry is the same as the line above it. Priors and MCMC details are given in Section \ref{sec: simDetails}} \centering
\begin{tabular}{rr|rr|rr|rr}
  \hline
  $\boldsymbol{\kappa}$ & \bf{Nominal} &      \multicolumn{2}{c|}{$\boldsymbol{\delta = 0.03}$}   &  \multicolumn{2}{c|}{$\boldsymbol{\delta = 0.02}$}   &  \multicolumn{2}{c}{$\boldsymbol{\delta = 0.01}$}    \tabularnewline

&
 & \textbf{Coverage} & \textbf{Mean } $\hat{p}$ & \textbf{Coverage} & \textbf{Mean } $\hat{p}$ & \textbf{Coverage} & \textbf{Mean } $\hat{p}$ \\ 
  \hline
 1 & 0.8 & 0.86 (0.05) & 38.48 (0.8) & 0.87 (0.06) & 45.65 (1.8) & 0.86 (0.05) & 55.20 (9.2) \\ 
&  0.9 & 0.94 (0.04) & ''& 0.94 (0.04) & '' & 0.94 (0.03) & '' \\ 
&  0.95 & 0.98 (0.02) & ''& 0.98 (0.02) & '' & 0.98 (0.02) & '' \\ \\
   2 &   0.8 & 0.82 (0.05) & 32.65 (0.9) & 0.80 (0.06) & 41.27 (1.2) & 0.84 (0.06) & 50.50 (1.9) \\ 
&  0.9 & 0.92 (0.04) & ''& 0.90 (0.05) & '' & 0.91 (0.04) & '' \\ 
& 0.95 & 0.96 (0.03) &'' & 0.95 (0.04) & '' & 0.95 (0.03) &'' \\ \\
 4 & 0.8 & 0.86 (0.06) & 28.38 (0.67) & 0.79 (0.05) & 36.67 (0.9) & 0.80 (0.07) & 48.45 (1.0) \\ 
&0.90 & 0.93 (0.04) & ''& 0.89  (0.04) & '' & 0.91 (0.04) & '' \\ 
& 0.95 & 0.97 (0.03) & ''& 0.94 (0.03) & ''& 0.97 (0.02) & '' \\ 
   \hline
\end{tabular}
\label{tab: variedDelta}
\end{table}

\subsection{Varying GSCM parameters}
\label{sec: variedShapes}

We also evaluate contour models that have mean values, $\boldsymbol{\mu}$,  that vary more slowly and more quickly  than in  {\it{Shape A}}.  These GSCMs are denoted as {\it{Shape B}} and {\it{Shape C}} respectively. Figure \ref{fig: shapesAll} shows sample contours and probability distributions for these shapes. Both models are defined to have $p = 50$ and $\kappa = 2.$ Exact values for  $\boldsymbol{\mu}$ and $\boldsymbol{\sigma}$ for ${\textit{Shape B}}$ and ${\textit{Shape C}}$  can be found in Supplement Section \ref{supp: simPars}. We report coverage results in Table \ref{tab: shapes}
for three $\alpha$-levels using  $N = 20 $ simulated observed samples and $\delta = 0.02$.

We find reasonably accurate coverage performance for all three shapes. The method performs slightly worse for \textit{Shape C} than for \textit{Shape A} and \textit{Shape B}. This performance difference is likely due to the difficulty of getting the pointed sections of \textit{Shape A} in the correct location if $\hat{p}$ is even slightly underestimated. This result indicates that for contours that look like \textit{Shape A}, lower $\delta$ values may be appropriate. Overall, the good performance across parameter settings indicates that a range of contours  can be well approximated by a GSCM.

\begin{table}
\caption{Mean coverage values for 40 simulations fitting the contour distribution for shapes A, B, and C.  In each simulation,  $N = 20$ simulated observed contours were sampled as training data and $M = 100$ evenly-spaced test lines were evaluated with $\theta^{*}_{1} = \pi/M$. Standard deviations across the test lines are given in parentheses. Priors and MCMC details are given in Section \ref{sec: simDetails}.  }
\centering
\begin{tabular}{r|rrr}
  \hline
 \textbf{Nominal}& \textbf{Shape A} & \textbf{Shape B}& \textbf{Shape C}\\ 
  \hline
0.8 & 0.79 (0.07) & 0.79 (0.06) & 0.87 (0.05) \\ 
  0.9 & 0.89 (0.05) & 0.89 (0.05) & 0.95 (0.03) \\ 
  0.95 & 0.95 (0.03) & 0.95 (0.04) & 0.98 (0.02) \\ 
   \hline
\end{tabular}
\label{tab: shapes}
\end{table}

\section{Modeling contours enclosing approximately star-shaped polygons}
\label{sec: approxStar}
A GSCM can also be applied to data where the polygons enclosed by the contours are not exactly star-shaped, but are approximately star-shaped. This section describes how to assess whether the the GSCM is appropriate given the observed data, and how the fitting procedure is altered if the observed contours enclose polygons that are not exactly star-shaped.

\subsection{Assessing appropriateness of GSCM}
\label{sec: starShapedMetric}

Two main assumptions must be met to apply the GSCM: the polygons enclosed by the contours must be approximately star-shaped and all contours should have at least one common point. The latter assumption is needed to define a starting point and can be trivially assessed. The former assumption can be assessed using metrics that describe how close an observed contour is to enclosing a polygon that is star-shaped. These metrics focus on the difference in the area between the polygon enclosed by the true contour and the polygon enclosed by the star-shaped representation of the contour.  If these differences are small for a set of observed contours, $\bold{S}$, then the GSCM can be applied.

We relax Definition \ref{def: starShapedApprox} to make precise how to approximate an arbitrary polygon with a star-shaped representation.  Two main differences between star-shaped polygons and arbitrary polygons are addressed in these new definitions: 1) an arbitrary polygon may not have a kernel and 2)  the contour line enclosing an arbitrary polygon may intersect with some of the lines in the line set multiple times.  The new star-shaped representation  definitions are:

\begin{definition}{Underestimated star-shaped representation, $\boldsymbol{\tilde{V}}_{u}(\boldsymbol{C}, \boldsymbol{\theta}, \boldsymbol{S})$: } \label{def: vUnder}Let  $\boldsymbol{\underline{S}}$ be a polygon described by ordered spatial points $\boldsymbol{S} = (s_{1}, \hdots, s_{n})$,  let $\boldsymbol{C} \in \boldsymbol{\underline{S}}$ be a starting point, let $\boldsymbol{\theta}$ be an arbitrary set of $p$ unique angles, and let $\boldsymbol{y} = (y_{1}, \hdots, y_{p})$ be a set of distances from $\boldsymbol{C}$ to the \textbf{closest} intersection point of the contour line $\boldsymbol{\overline{S}}$ and each line $\ell_{i}$ in the line set  $\mathcal{L}( \boldsymbol{C}, \boldsymbol{\theta})$. Then, the star-shaped representation of the contour, $\boldsymbol{V}(\boldsymbol{C}, \boldsymbol{\theta}, \boldsymbol{y})$,  is the underestimated star-shaped representation,  $\boldsymbol{\tilde{V}}_{u}(\boldsymbol{C}, \boldsymbol{\theta}, \boldsymbol{S})$.  
\end{definition}

\begin{definition}{Overestimated star-shaped  representation, $\boldsymbol{\tilde{V}}_{o}(\boldsymbol{C}, \boldsymbol{\theta}, \boldsymbol{S})$:} \label{def: vOver} Let  $\boldsymbol{\underline{S}}$ be a polygon described by ordered spatial points $\boldsymbol{S} = (s_{1}, \hdots, s_{n})$,  let $\boldsymbol{C} \in \boldsymbol{\underline{S}}$ be a starting point, let $\boldsymbol{\theta}$ be an arbitrary set of $p$ unique angles, and let $\boldsymbol{y} = (y_{1}, \hdots, y_{p})$ be the set of distances from $\boldsymbol{C}$ to the \textbf{farthest} intersection point of the contour line $\boldsymbol{\overline{S}}$ and each line $\ell_{i}$ in the line set  $\mathcal{L}( \boldsymbol{C}, \boldsymbol{\theta})$. Then, the star-shaped representation of the contour, $\boldsymbol{V}(\boldsymbol{C}, \boldsymbol{\theta}, \boldsymbol{y})$,  is the overestimated star-shaped representation,  $\boldsymbol{\tilde{V}}_{o}(\boldsymbol{C}, \boldsymbol{\theta}, \boldsymbol{S})$.  
\end{definition}

\begin{figure}[ht!]
\centering
\includegraphics[width=.7\textwidth]{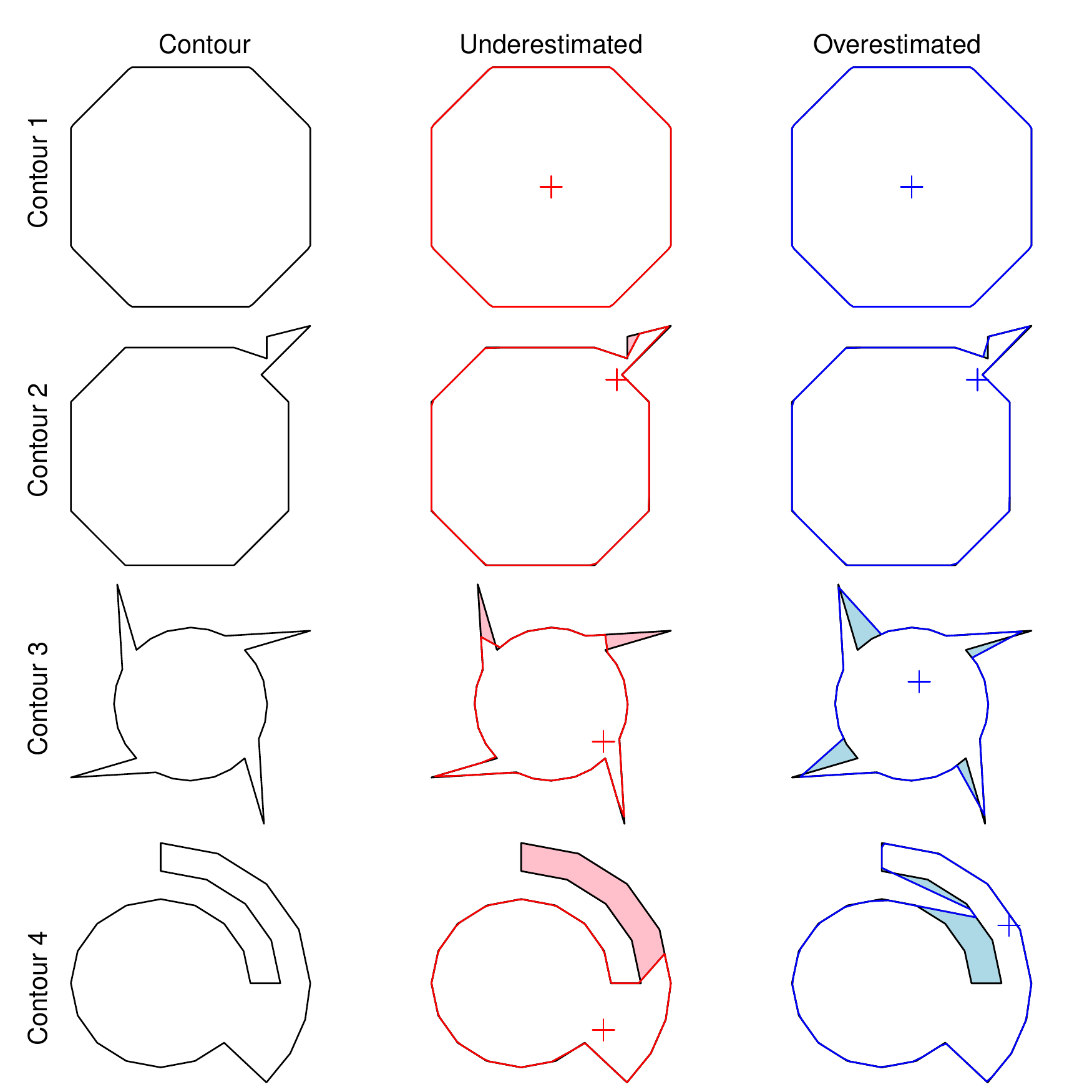}
\caption{  Four contours (left) with their underestimated star-shaped approximations, $\boldsymbol{\tilde{V}}_{u}(\boldsymbol{\hat{C}}_{u}, \boldsymbol{\theta}, \boldsymbol{S})$, (red, center), and their overestimated star-shaped approximations,  $\boldsymbol{\tilde{V}}_{o}(\boldsymbol{\hat{C}}_{o}, \boldsymbol{\theta}, \boldsymbol{S})$ (blue, right).  The polygon in the top row is star-shaped,  and the other three polygons are not.  Pink and blue sections in the center and right panel  show the respective differing areas, $A_{u}(\boldsymbol{\hat{C}}_{u}, \boldsymbol{\theta}, \boldsymbol{S})$ and $A_{o}(\boldsymbol{\hat{C}}_{o}, \boldsymbol{\theta}, \boldsymbol{S})$. The red crosses  in the central panel and the blue crosses in the right panel denote the estimated starting points, $\boldsymbol{\hat{C}}_{u}$ and $\boldsymbol{\hat{C}}_{o}$.  The vector $\boldsymbol{\theta}$ contains 200 elements spaced evenly in the interval $[0, 2\pi]$.}
\label{fig: starShapedMetric}
\end{figure}

The names of these representations highlight that these polygons generally under- or overestimate the area contained within the true polygon. For contours that enclose star-shaped polygons, if $\boldsymbol{C} \in \mathcal{K}(\boldsymbol{\underline{S}})$, only one intersection is found between the contour line and all lines in the line set. So,  $\boldsymbol{\tilde{V}}_{o}(\boldsymbol{C}, \boldsymbol{\theta}, \boldsymbol{S}) = \boldsymbol{\tilde{V}}_{u}(\boldsymbol{C}, \boldsymbol{\theta}, \boldsymbol{S}) = \boldsymbol{\tilde{V}}(\boldsymbol{C}, \boldsymbol{\theta}, \boldsymbol{S})$. For notational convenience, let $\boldsymbol{\tilde{V}}_{u} = \boldsymbol{\tilde{V}}_{u}(\boldsymbol{C}, \boldsymbol{\theta}, \boldsymbol{S})$ and $\boldsymbol{\tilde{V}}_{o} = \boldsymbol{\tilde{V}}_{o}(\boldsymbol{C}, \boldsymbol{\theta}, \boldsymbol{S})$.  Then the sets that differ between the true contour and the under- and overestimated star-shaped representations are 
\begin{align}
A_{u} (\boldsymbol{C}, \boldsymbol{\theta}, \boldsymbol{S}) &:= \{ (\underline{\boldsymbol{S}}^{c} \cap \underline{\boldsymbol{\tilde{V}}_{u}})  \cup (\underline{\boldsymbol{S}} \cap \underline{\boldsymbol{\tilde{V}}_{u}}^{c})\} , \\
A_{o}(\boldsymbol{C}, \boldsymbol{\theta}, \boldsymbol{S})  & := \{ (\underline{\boldsymbol{S}}^{c} \cap \underline{\boldsymbol{\tilde{V}}_{o}})  \cup (\underline{\boldsymbol{S}} \cap \underline{\boldsymbol{\tilde{V}}_{o}}^{c})\} ,
\end{align}
where the superscript $c$ denotes the complement of the set. Let $|A_{u} (\boldsymbol{C}, \boldsymbol{\theta}, \boldsymbol{S}) |$ and $|A_{o}(\boldsymbol{C}, \boldsymbol{\theta}, \boldsymbol{S}) |$  be the area contained within these sets. Figure \ref{fig: starShapedMetric}  illustrates the quantities described in this section for four contours. 
The difference in area is zero only if the polygons are star-shaped. More precisely:

\begin{theorem}
\label{theorem: posArea}
For any polygon $\boldsymbol{\underline{S}}$  that is not star-shaped  $|A_{u} (\boldsymbol{C}, \boldsymbol{\theta}, \boldsymbol{S})| > 0$ and $|A_{o} (\boldsymbol{C}, \boldsymbol{\theta}, \boldsymbol{S})|> 0$ for any $\boldsymbol{C}$ and $\boldsymbol{\theta}$. (Proof in Appendix \ref{app: proofs}.)
\end{theorem}

For easier comparison of these differences in area across sets of polygons of different sizes, these areas can be expressed as percentages of the mean total area of the polygons.  Table 4 reports the difference in area for the contours  in Figure \ref{fig: starShapedMetric}, and their star-shaped representations as a percentage of the total area of the polygon.

\subsection{Fitting GSCMs to approximately star-shaped polygons}
\label{sec: fitModelMisspec}
The approach to fitting in Section \ref{sec: fitting} needs to be altered slightly for contours that enclose regions that are only approximately star-shaped contours. The values of $\boldsymbol{y}$ need to be computed using  the under- or overestimated star-shaped approximation as given in Definitions \ref{def: vUnder} and \ref{def: vOver}.  Whether to use the under- or overestimated star-shaped representation depends on the application. In some cases, asymmetric risks may motivate selecting a model that generally over- or underestimates the area within the polygon. Otherwise, both $|A_{u} (\boldsymbol{C}, \boldsymbol{\theta}, \boldsymbol{S}) |$ and $|A_{o}(\boldsymbol{C}, \boldsymbol{\theta}, \boldsymbol{S}) |$ can be computed for the set of observed contours $\boldsymbol{S}$ and whichever representation results in less difference in area can be selected.  Fitting the posterior proceeds as in Section \ref{sec: compPost}.  Finding probabilities and credible intervals proceeds as in Section \ref{sec: probsCredInts}.

 We find $\boldsymbol{\hat{C}}$ and $\boldsymbol{\hat{\theta}}$ using nearly the same algorithm as in Section \ref{sec: determineCAndP}, except that we update the star-shaped representation  in Equation \ref{eq: CHatOptD} and Equation \ref{eq: redP} to be  the under- or overestimated star-shaped representation. Specifically,  we  replace Equation \ref{eq: CHatOptD} with 
\begin{align}
\label{eq: findCEstU}
\boldsymbol{\hat{C}}_{u} = \underset{\boldsymbol{C} \in \boldsymbol{\mathcal{D}}}{\text{argmin}} \left \{ f ( |A_{u} (\boldsymbol{C}, \boldsymbol{\theta}, \boldsymbol{S}_{1})|, \hdots, |A_{u} (\boldsymbol{C}, \boldsymbol{\theta}, \boldsymbol{S}_{N})|)\right\} 
\end{align}

\begin{table}
\caption{The differing area for the under- and overestimated star-shaped approximations, $|A_{u}(\boldsymbol{\hat{C}}_{u}, \boldsymbol{\theta}, \boldsymbol{S})|$  and  $|A_{o}(\boldsymbol{\hat{C}}_{o}, \boldsymbol{\theta}, \boldsymbol{S})|$, for the contours in  Figure \ref{fig: starShapedMetric}. Differences in area are computed numerically and expressed as a percentage of the total area of the polygon. The vector $\boldsymbol{\theta}$ contains 200 elements spaced evenly in the interval $[0, 2\pi]$.   }
\centering
\begin{tabular}{r|rr}
\hline
 &\textbf{Underestimated} & \textbf{Overestimated} \\ 
  \hline
\bf{Contour 1} & 0.00 & 0.00 \\ 
  \bf{Contour 2} & 0.43 & 0.24 \\ 
  \bf{Contour 3} & 4.44 & 9.78 \\ 
  \bf{Contour 4} & 15.65 & 8.46 \\ 
   \hline
\end{tabular}
\label{tab: differingAreas}
\end{table}

\noindent for the underestimated representation and 
\begin{align}
\label{eq: findCEstO}
\boldsymbol{\hat{C}}_{o} = \underset{\boldsymbol{C} \in \boldsymbol{\mathcal{D}}}{\text{argmin}} \left\{ f ( |A_{u} (\boldsymbol{C}, \boldsymbol{\theta}, \boldsymbol{S}_{1})|, \hdots, |A_{u} (\boldsymbol{C}, \boldsymbol{\theta}, \boldsymbol{S}_{N})|) \right\}
\end{align}
for the overestimated representation.  The values $\boldsymbol{\hat{C}}_{u}$ and $\boldsymbol{\hat{C}_{o}}$ denote the estimated starting points for the two approximations respectively. As before,  $f$ is the mean function.  Like in Section \ref{sec: determineCAndP}, we evaluate these functions of the area difference at a grid of possible locations, $\boldsymbol{\mathcal{D}}$. Unlike in Section \ref{sec: determineCAndP} we cannot use the estimated intersection kernel to reduce the size of $\boldsymbol{\mathcal{D}}$,  since the polygons are not exactly star-shaped. However, we note that the starting point must be common to all observed polygons, so we constrain $\boldsymbol{\mathcal{D}}$ to only cover areas in the intersection of all observed polygons, $\boldsymbol{\underline{S}}_{1} \cap \hdots \cap \boldsymbol{\underline{S}}_{N}$. Equation \ref{eq: redP} is also slightly changed so that
\begin{align}
\label{eq: redP2}
f (|A_{u} (\boldsymbol{C}, \boldsymbol{\hat{\theta}}, \boldsymbol{S}_{1})|, \hdots, |A_{u} (\boldsymbol{C}, \boldsymbol{\hat{\theta}}, \boldsymbol{S}_{N})| ) < \frac{\delta}{N} \sum_{i=1}^{N} |\boldsymbol{\underline{S}_{i}}|
\end{align}
for the underestimated representation and
\begin{align}
f (|A_{o} (\boldsymbol{C}, \boldsymbol{\hat{\theta}}, \boldsymbol{S}_{1})|, \hdots, |A_{o} (\boldsymbol{C}, \boldsymbol{\hat{\theta}}, \boldsymbol{S}_{N})| ) <   \frac{\delta}{N}\sum_{i=1}^{N} |\boldsymbol{\underline{S}_{i}}|
\end{align}
for the overestimated representation.  The value $f$ still represents the mean function and  $\delta$ is still a proportion.

\subsection{Coverage metric for approximately star-shaped polygons }
 This metric introduced in Section \ref{sec: starShapedMetric} is essentially the same when applied to approximately star-shaped polygons.  However, $R_{i, k}$, the intersection of an observed contour $\boldsymbol{S}_{i}$ and line $\ell_{k}$, may now contain multiple points since polygon $\boldsymbol{\underline{S}}_{i}$ is only approximately star-shaped.  A single point of intersection will still be more common when polygons are approximately star-shaped.   Aside from this change, the definition of $W_{i, k} = \mathbbm{1}[R_{i, k} \in I_{1 - \alpha, k}]$  remains the same when $R_{i, k}$ is composed of multiple points. So, all subsequent definitions and properties in Section  \ref{sec: starShapedMetric} hold as well.

\begin{figure}
\centering
\includegraphics[width=.5\textwidth]{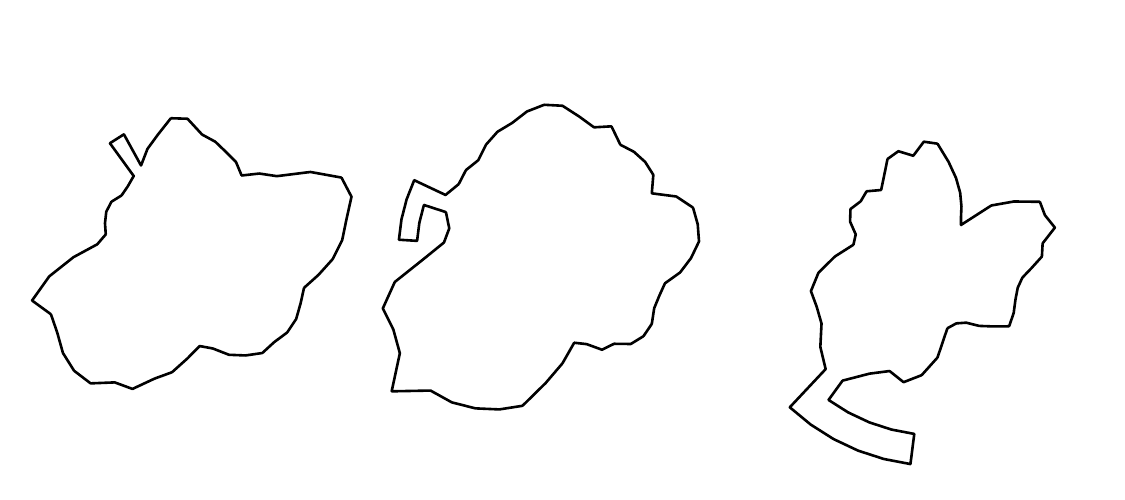}
\caption{Three generated contours that enclose polygons that are only approximately star-shaped. From left to right, contours have increasing area that cannot be described with a star-shaped representation. } 
\label{fig: notStarShaped}
\end{figure}

\subsection{Simulation study of contours enclosing approximately star-shaped polygons }
\label{sec: ApproxStarShape}

In these simulations, we assess GSCM performance for contours that enclose polygons that are only  approximately star-shaped. We simulate contours that vary systematically in how much the polygons they enclose differ from being star-shaped.  Figure \ref{fig: notStarShaped} shows examples of the types of contours we will evaluate. 

To obtain these contours, we first generate polygons that are star-shaped from a GSCM. We then append sections to these polygons that cause the polygons to no longer be star-shaped. The appended sections  loop back around the outside of the initial polygon over some  number of lines in the initial line set.   The number of initial lines looped back over are selected randomly from some uniform distribution. Appended sections that loop around a larger number of lines are longer than appended sections that loop over a smaller number of lines.  Longer appended sections result in more area that cannot be described with a star-shaped representation than shorter appended sections. How close the appended section is to the initial star-shaped polygon and the width of the appended section are selected randomly from uniform distributions. 

We consider 40 evaluation runs for three different uniform distributions for the number of initial lines that the appended section loops back over. The initial GSCM is set to be \textit{Shape A} with $\kappa = 2$ and  $p = 50$. In each evaluation run, the GSCM is fit to $N = 20$ simulated contours. Rather than fixing $\delta$ to determine $\hat{p}$ and $\boldsymbol{\hat{C}}$ in these simulations, we set $\hat{p} = p = 50$. We then estimate $\boldsymbol{\hat{C}}$ conditional on $\hat{p}$. This choice simplifies the interpretation of our results. (Larger $\delta$'s would be needed as the area differing from a star-shaped polygon increases. So, it would be difficult to distinguish whether performance differences were due to changes in $\hat{p}$ or changes in how close to star-shaped the polygons are.) We initially run these simulations using a random location for the appended section and then repeat these simulations with a fixed location for the appended section. Results are reported in Table \ref{tab: misspec}. 

\begin{figure}
\centering
\includegraphics[width=.7\textwidth]{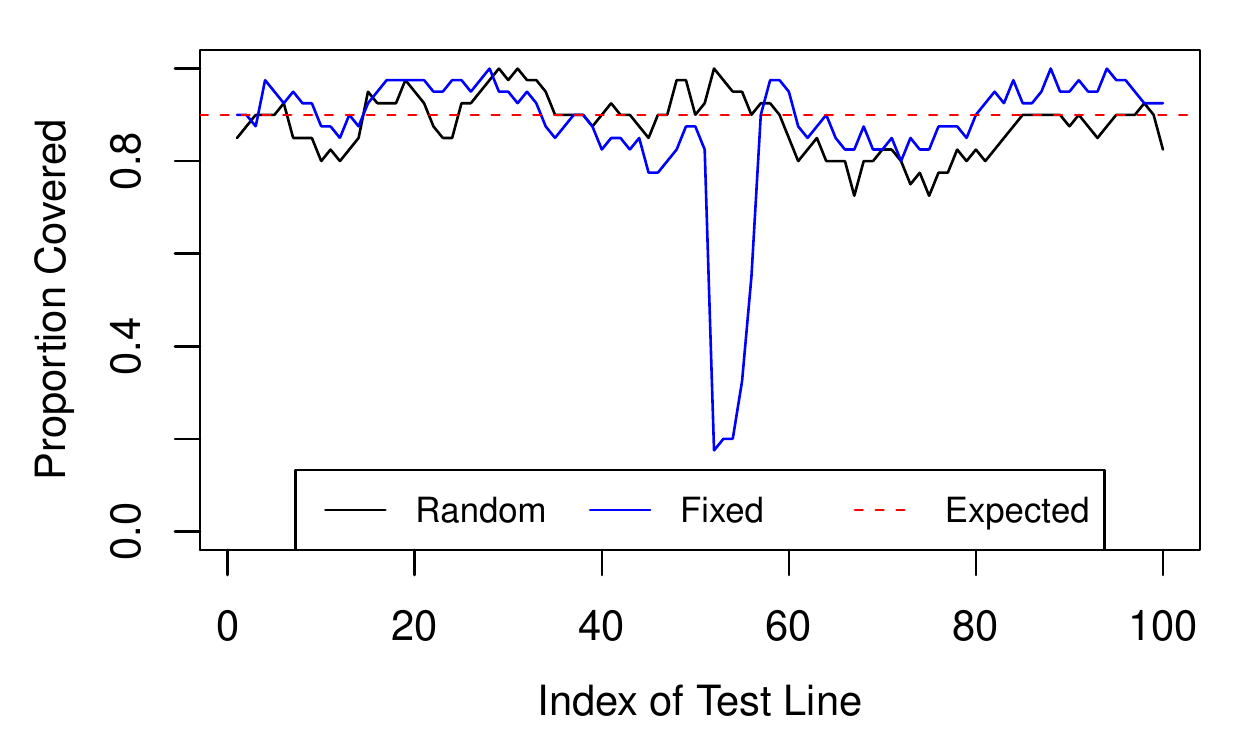}
\caption{Proportion of lines covered out of 40 evaluation runs plotted against the index of each of $M = 100$ evenly-spaced test lines for the 90$\%$ credible interval for the contours enclosing approximately star-shaped polygons. The black line corresponds to when the appended sections are added to a random location and the blue lines corresponds to when the appended sections are added to a fixed location. The number of initial lines looped back over are selected randomly from a Uniform (4, 5) distribution. Nominal coverage is in red. Priors and MCMC details are given in Section \ref{sec: simDetails}.  } 
\label{fig: misspecTransect}
\end{figure}

With a random location for the appended section, we find that applying GSCMs still results in reasonable coverage for the 90$\%$ and $95\%$ credible intervals.  For the $80\%$ credible interval, performance is moderately degraded. Interestingly, average performance does not seem to be correlated with how large the appended section to the contour is. When an appended section is added to a fixed location, we find that the mean coverage across the test lines  is relatively accurate; however, the standard deviation is quite high, suggesting that coverage is actually poor in some parts of the contours. Figure \ref{fig: misspecTransect} illustrates this variability in performance for the case with the number of initial lines looped over distributed Uniform(4, 5). We plot  the proportion of evaluation runs covered by the $90\%$ credible intervals for each test line individually.     

The location of the fixed appended section is under-covered. In contrast, no obvious patterns are seen when the location of the appended section is random. These results indicate that contours  that enclose polygons that modestly differ from star-shaped contours can be modeled with GSCMs.  However, if areas differing from the star-shaped representation occur in the same location repeatedly, additional modeling of these areas may be needed to avoid systematic errors in coverage.

\begin{table}
\caption{Mean coverage values for 40 simulations of fitting the contour distribution for approximately star-shaped data.  The number of initial lines for the appended section to loop back over are selected randomly from a Uniform(a, b) distribution.  In each simulation,  $N = 20$ simulated observed contours were sampled as training data and $M = 100$ evenly-spaced test lines with $\theta_{1}^{*} = \pi/M$  were evaluated. The appended sections are located in a random position in the first three cases and a fixed location in the second three cases. Standard deviations across the test lines are given in parentheses. Priors and MCMC details are given in Section \ref{sec: simDetails}}
\centering
\begin{tabular}{r|rrrrrr}
  \hline
&\textbf{Nominal} & \textbf{Unif}$\boldsymbol{(0, 1)}$ & $\textbf{Unif}\boldsymbol{(2, 3)}$ & \textbf{Unif}$\boldsymbol{(4, 5)}$\\
\hline
  \textbf{Random Location:}& 0.8 & 0.90 (0.04) & 0.88 (0.07) & 0.90 (0.05) \\ 
 & 0.9 & 0.96 (0.02) & 0.94 (0.04) & 0.93 (0.04) \\ 
 & 0.95 & 0.98 (0.02) & 0.97 (0.02) & 0.95 (0.03) \\ \\
     \textbf{Fixed Location:}  & 0.8 & 0.79 (0.08) & 0.77 (0.12) & 0.77 (0.16) \\ 
 & 0.9 & 0.88 (0.06) & 0.87 (0.10) & 0.87 (0.15) \\ 
 & 0.95 & 0.94 (0.04) & 0.93 (0.07) & 0.93 (0.13) \\ 
   \hline
   \hline
\end{tabular}
\label{tab: misspec}
\end{table}

\section{Example: September Arctic sea ice edge contour}
\label{sec: example}

\begin{figure}
\centering
\includegraphics[width=.7\textwidth]{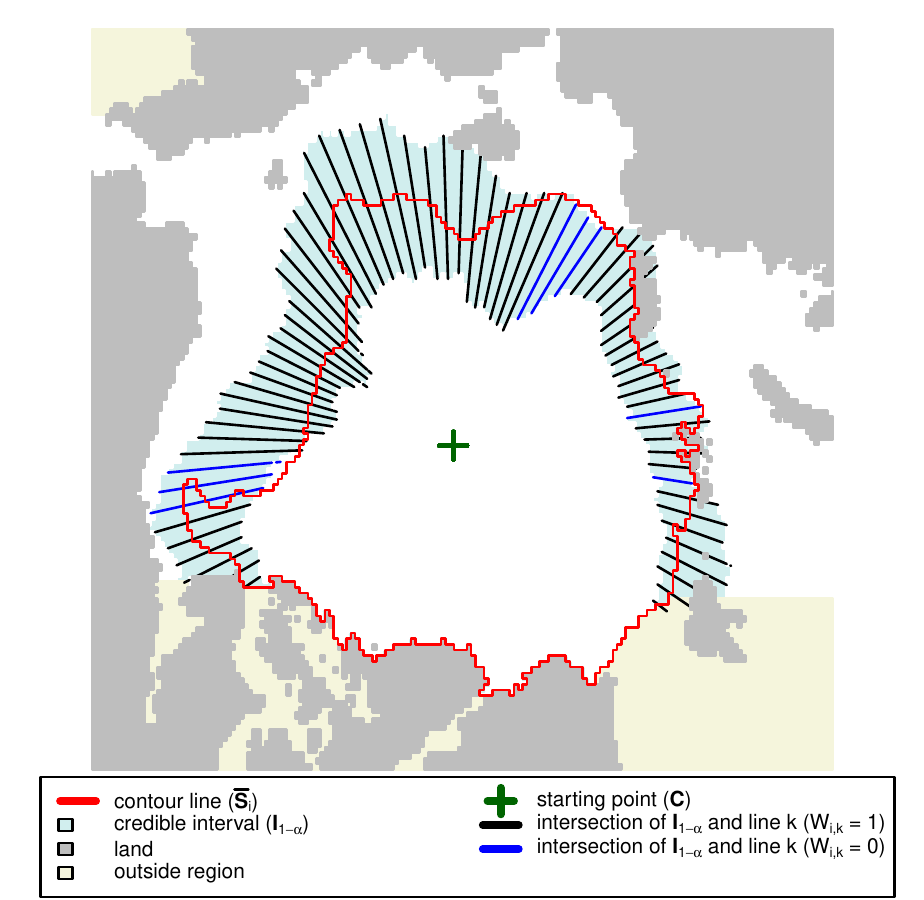}
\caption{The September 2017 sea ice edge contour for the central Arctic region (red) with an 80$\%$ credible region fitted from the GSCM with data from 2008-2016. Line segments, $I_{0.8, k}$,  corresponding to the intersection of the $\boldsymbol{I}_{0.8}$ credible region and line $\ell_{k}$  are colored black when they cover the contour and blue otherwise. The starting point of the evaluation is denoted by a green cross sign. } 
\label{fig: seaIceEx}
\end{figure}

We now return to the motivating example of the Arctic sea ice edge. Modeling the distribution of the sea ice edge is useful for understanding how sea ice is changing over time.  Also, predicting the sea ice edge is helpful for maritime planning, since traversing over sea ice is expensive in time and fuel. We focus on modeling the sea ice edge in September, the month when the ice-covered area in the Arctic is at its annual minimum. Maritime vessel traffic is typically highest in September because the largest portion of the Arctic ocean is not ice-covered. 

Arctic sea ice edge contours observed in previous Septembers provide some information about what types of sea ice edges might be expected in the near future.  However, ice edges from past decades are unlikely to be seen again, since, with climate change, the Arctic has seen a rapid reduction in the area covered by sea ice  \citep{Comiso2008, Stroeve2012}. September sea ice varies substantially from year to year, so recent observations cover only a small fraction of the physically plausible ice edges that could occur. Therefore, we need a model for the sea ice edge to make statements about how probable sea ice edges will be in the near future and/or to describe plausible sea ice edges that could have been observed in past years' conditions. The GSCM can be applied to generate plausible sea ice edge contours and corresponding credible intervals for the ice edge contour.  

To assess the sea ice edges generated by the GSCM, we perform a leave-one-out cross validation experiment on the September sea ice edge contour in a region in the central Arctic  for ten recent years. Data on sea ice are obtained from a monthly-average observational product produced by the National Aeronautics and Space Administration satellites Nimbus-7 SMMR and DMSP SSM/I-SSMIS and downloaded from the National Snow and Ice Data Center \citep{Comiso2000}. Each grid box in this data product has an associated concentration, or percentage of ice-covered area. To form the sea ice edge, we categorize all grid boxes with at least 15$\%$ concentration as containing sea ice and all grid boxes with less than 15$\%$ concentration as open water. This thresholding is often used by sea ice researchers, because the transition from complete sea ice cover to open water occurs over a narrow spatial range and satellite estimates for the low concentrations are not considered reliable.

We assume for this analysis that the ice edge contours observed over the different years are independent and that the contour distribution of the ice edge is stationary over time.  While this stationarity assumption is not strictly true given climate change, for a time period of ten years  the effects of the climate change trend are small relative to year-to-year variability. For each year $j$ from 2008 to 2017, we fit the GSCM using the contours observed in the other nine years. We  then try to ``predict" the distribution of possible ice edges that would have been plausible in year $j$. Data have been rescaled as described in Section \ref{sec: rescaleData} with  $\epsilon = 0.1$. We exclude from fitting a section of the ice edge contours that always borders land, since there is no variability in the associated $\boldsymbol{y}$ values. We use the overestimated approximation, since for some maritime planning applications, predicting too much sea ice may be less dangerous than predicting too little sea ice.

As shown in Section \ref{sec: nObs}, nine years of data may not be sufficient for fitting a GSCM well.  In this context, though, we have considerable information that we can incorporate into the prior to augment the observations. In particular, the contour in the area we are trying to predict is bounded primarily by land and also by the boundary around the region. So, the lines in the line set $\mathcal{L} = (\ell_{1}, \hdots, \ell_{p})$ are also bounded. We leverage their fixed maximal lengths in setting the priors. Specifically, we let  $\mu_{0, i} = 0.5\ell_{i}$ and $\beta_{0, \sigma_{i}} = (\ell_{i}/2)/\Phi^{-1}(.995)$. The latter corresponds to the standard deviation of  a normal distribution with $99\%$ of its mass falling in the interval  $(0, ||\ell_{i}||)$ where $||\ell_{i}||$ is the length of $\ell_{i}$.  We also  set $\Lambda_{0}$ be a diagonal matrix with $0.5^{2}$ on the diagonal and  $\beta_{\kappa_{0}} = 10$. We use  MCMC  for fitting with 50,000 iterations, of which 25,000 are omitted as burn-in. 

 We evaluate coverage  as in Section \ref{sec: coverageMetric} using the approximately optimal starting point  selected using the method in Section \ref{sec: fitModelMisspec} with $\delta = 0.05$ and $p = 100$.  We exclude from evaluation test lines that always border land and have no variability. This leaves 71 test lines. To illustrate this coverage evaluation, we plot in Figure \ref{fig: seaIceEx} the 80$\%$ credible interval estimated for the 2017 sea ice edge fit with data from the other nine years. As in Section \ref{sec: coverageMetric} lines extending outward from the estimated starting point are colored blue if they do not cover the left-out contour and black otherwise. 
Table \ref{tab: seaIceTable} reports the mean coverage of the credible intervals averaged over the ten years and 71 testing lines. Three nominal $\alpha$-levels are considered. The observed and nominal coverage values are similar, suggesting that the GSCM  has appropriate coverage. More broadly, we have shown that the GSCM has potential for generating and describing contours of scientific interest.

\begin{table}
\caption{Mean coverage for a leave-one-out cross-validation study for the 2008-2017 September sea ice edge. $M = 71$ evenly-spaced test lines were evaluated with $\theta^{*}_{1} =\pi/M$. Standard deviations across the test lines are given in parentheses. }
\centering
\begin{tabular}{rrrr}
  \hline
\bf{Nominal} & \bf{Mean }  \\ 
  \hline
0.80 & 0.75 (0.12)\\ 
0.90 & 0.87 (0.08) \\ 
0.95 & 0.92 (0.07) \\ 
   \hline
\end{tabular}
\label{tab: seaIceTable}
\end{table}

\section{Discussion}
\label{sec: discussion}

We have introduced the GSCM for modeling  contours that enclose polygons that are star-shaped or approximately star-shaped.  Simulation studies illustrated how  GSCMs provide accurate coverage in different scenarios. Analysis of September Arctic sea ice also showed how GSCMs can be useful for applied problems.  We conclude this paper with a discussion of how GSCMs relate to other contour models and directions for future research.

\subsection{Other approaches}
A large body of research addresses contours in other contexts.  GSCMs are applied when multiple contour boundaries are directly observed and the distribution of possible contours is the primary object of inference. Other methods for contours and boundaries may be appropriate for other applications. 

Much of the existing research on boundaries relates to level exceedances, also called excursions. The classic level exceedance problem refers to inferring the contour enclosing regions where some latent process exceeds a certain level $u$. Inference is based on measurements taken at various spatial points on a random spatial field. Early work by  \citet{Polfeldt1999}  considers how to make statements about the accuracy of contour maps in this context.  \citet{Lindgren1995} first define contour uncertainty regions using unions of crossing intervals, or line sections where transitions from below and above $u$ occur. More recently, \citet{Bolin2015} introduce a method for inferring exceedance levels with irregularly-spatial measurements when the latent spatial field is Gaussian. Their approach provides a way to make global statements about the uncertainty of the full contour. \citet{Bolin2017} then extend this method to estimate the uncertainty of multiple contours to produce contour maps with appropriate uncertainty estimates. Both methods leverage Integrated Nested Laplace Approximations for efficient computation and can be implemented with the {\tt{excursions}} R package \citep{Bolin2018}. 

\citet{French2014} provides an alternate simulation-based method for  making global statements about the location of the contour. Methods for identifying the exceedance region are also explored from both Bayesian and frequentist perspectives \citep{French2013, French2016}.  Level exceedance methods and GSCMs both focus on inferring contours and their uncertainty. However, in the former, boundaries and their uncertainty are inferred with measurements of a continuous process made at spatially-referenced points while in the latter distributions of plausible contours are inferred from direct observations of the contour boundaries themselves. 

Wombling methods also focus on spatial boundaries.  First considered by Womble \citep{Womble1951}, these methods typically apply bilinear interpolation  to spatially-referenced data points. The gradients of the interpolated functions are used to infer boundaries.  \citet{Jacquez2000} summarizes early research on primarily deterministic methods for identifying these boundaries. Principled Bayesian statistical methods have since been developed  \citep{Banerjee2006, Gelfand2015} and recent research has introduced Wombling methods for areal data \citep{Lu2005, Li2015} and point processes \citep{Liang2009}. Wombling boundaries are inferred from spatially-referenced or areal data. The proposed GSCM  differs from Wombling techniques, since it is targeted to be applied to repeated directly-observed contour boundaries.

Statistical shape analysis \citep[e.g.][]{Dryden2016, Srivastava2016} describes features and variation around boundaries. Shapes typically  have consistent and definable features. In these types of applications, location and rotational effects are often ignored in describing the distribution around the shape.  As an example, shape analysis is often applied to biological imaging research. Deformable templates were developed to describe distributions around shapes with definable features such as the parts of a hand \citep{Amit1991, Grenander1993, Grenander1998}.  The applications motivating GSCMs  differ from the applications motivating shape analysis  in that  the physical location of the boundary is of interest and no features are present.  

Many image analysis methods have been developed to  segment images or identify edges. Techniques such as mathematical morphology \citep{Haralick1987, Lee1987}, watershed segmentation \citep[e.g.][]{Gauch1999}, and more recently deep learning are applied to identify or sharpen the uncertainty of a single observed boundary in an image. The goals of these methods again differ from the goal of GSCM to  define variability over multiple observations of boundaries.

Another alternative to GSCMs would be to model directly whether the points on a lattice are inside or outside a contour boundary.  However, methods for modeling binary data on a lattice such as the autologistic \citep{Besag1974}, centered autologistic \citep{Caragea2009}, and the spatial generalized linear mixed model  \citep{Besag1991, Diggle1998, Hughes2013} are not structured to guarantee that all the grid boxes inside the contour form in a contiguous section. Hence, these methods are not designed directly for modeling contour boundaries.

\subsection{Fractal contours and GSCMs}
\label{sec: fracGSCMs}

\begin{figure}[h!]
\centering
\includegraphics[height = .55\textheight]{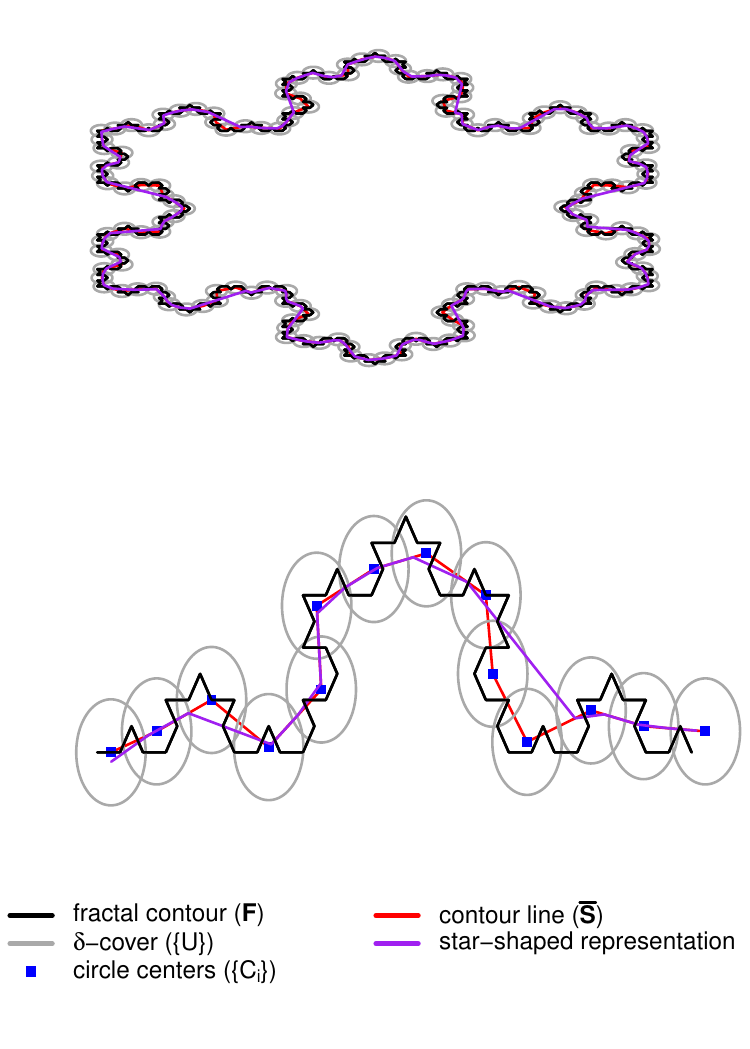}
\caption{Top: A visualization of a fractal contour, $\boldsymbol{F}$, with several self-similar layers shown (black line), a $\delta$-cover of $\boldsymbol{F}$ (grey circles), and a line connecting the starting points of the $\delta$-cover (red line). Bottom: Magnified section of the top figure with circle centers, $\boldsymbol{S}$, (blue squares), added.  Note that the $\delta$-cover shown is just a visualization and may not contain $N_{\delta}(F)$ elements. } 
\label{fig: fractal}
\end{figure}

We have treated contours as connected sequences of points, but  many contours have fractal-like properties. We now discuss how a fractal contour could be converted to a connected sequence of points for modeling with a GSCM.  A true fractal contour, represented by a set $F$, could be approximately represented by a smaller set of points $\boldsymbol{S}$.   In a  2-dimensional Euclidean space, $\mathbb{R}^{2}$, consider a countable or finite set $\{U \}$ of circles of radius $\delta$.  We say $\{U \}$ covers $F$ if $F \subset \cup_{i=1}^{\infty}U_{i}$ and we refer to the set $\{U \}$ as a $\delta$-cover of $F$. The value $N_{\delta}(F)$ is the smallest number of circles  of radius  $\delta$ that could be used to cover $F$ \citep[][pp. 27-28]{Falconer2004}.  Let $\{U^{*}\}$ denote one covering that contains $N_{\delta}(F)$ circles.  Since the contour $F$ is assumed to be finite, the number of circles in $\{U^{*} \}$ will also be finite.  In  Figure \ref{fig: fractal}, we plot a $\delta$-cover over a visualization of a fractal contour $F$ with several finite self-similar layers. The $\delta$-cover plotted is for visualization and may not contain exactly $N_{\delta}(F)$ circles.

 We can define the elements in the sequence of points forming the contour, $\boldsymbol{S}$,  to be the starting points,  $\{M^{*}\}$, of the circles in $\{U^{*}\}$.  The starting points should be arranged in the order in which they would be touched if one were to trace over the fractal contour line, $F.$  Since the contour is a closed loop, where and in what direction to start tracing the fractal contour, $F$, only affects the indexing of the starting points and not the contour line formed by connecting these points. With this procedure the distance from any point $F$ to a point in $\boldsymbol{S}$ is no more than $\delta$. In general there are multiple $\delta$-covers that contains $N_{\delta}(F)$ circles; therefore,  a criterion needs to be specified as to which set $\{U^{*}\}$ is used to define $\boldsymbol{S}$. For example, the set $\{U^{*}\}$  could be selected to be the $\delta$-cover with $N_{\delta}(F)$ circles that has the circle center closest to the highest $x-$ and highest $y-$ coordinate in the domain, i.e. the top right corner of the domain. 

Once $\boldsymbol{S}$ has been obtained, GSCMs can be used. In particular, the under- or overestimated star-shaped representation to any fractal contour can be found for $\boldsymbol{S}$ as described in Definitions \ref{def: vUnder} and \ref{def: vOver}. A star-shaped representation can represent the area contained within some fractal contours fairly well. For example, the visualized $F$ and the overestimated star-shaped representation for $\boldsymbol{S}$ obtained using the visualized $\delta$-cover are similar. The differing area represents only 7.4\% of the total area of the visualized fractal contour when $p = 200$ lines are used in the line set.

\subsection{Limitations and extensions}
A limitation of GSCMs is that they can be applied only to contours that enclose polygons that are star-shaped or approximately star-shaped. 
Using multiple starting points would be one way to relax this assumption.
From each starting point a different star-shaped contour could be generated. These individual contours could be combined to produce an overall contour that encloses a polygon that is not star-shaped. Another promising direction for relaxing the star-shaped assumption is to develop a method for appending sections to an initial contour that encloses star-shaped polygons. Extending the way contours were generated in the simulation in Section \ref{sec: ApproxStarShape} might provide an initial approach to do this. 
 
Another limitation is that  GSCMs  assume all generated line lengths $\boldsymbol{y}$ are drawn from a Gaussian distribution. While GSCMs can cover a range of shapes, in some applications the observed $y$ may be skewed, bounded above, or otherwise differ from being distributed approximately normally. Therefore, exploring alternatives to GSCMs that allow for more flexible distributions of $\boldsymbol{y}$ would be valuable. 

Overall, though, GSCMs provide a promising avenue for modeling data composed of multiple observed contours. Representing contours with sequences of points, which are of lower dimension than spatial fields, could enable more detailed and efficient modeling of contour boundaries.

%\begin{figure}
%\centering
%\includegraphics[scale=.9]{./figure/toy/particles.pdf}
%\includegraphics[scale=.9]{./figure/toy/neighbourhood.pdf}
%\caption{Illustration of one AIMM increment for the target $\pi_1$. Top: states $X_1,\ldots,X_n$ of the %Markov chain, proposed new state $\tilde{X}_{n+1}$ activating the increment process (\ie satisfying $W_n(\tilde{X}_{n+1})\geq \Wst$) and neighborhood of $\tilde{X}_{n+1}$. Bottom: target $\pi_1$, defensive kernel $Q_0$, first increment $\phi_1$ and updated kernel $Q_1$ plotted in log-lin scale.}
%\label{fig:ill1}
%\end{figure}

%\vspace{-.5cm}
%\section*{Acknowledgements}

\clearpage
\bibliographystyle{apalike}
\bibliography{references_contourTheory}

\newpage

\appendix

\section{Appendix: Proofs} 
\label{app: proofs}

\subsection{Proof of Theorem \ref{theorem: existsV}}
Let $\boldsymbol{\theta} = (\theta_{1}, \hdots, \theta_{p})$ with $\theta_{i} < \theta_{i + 1}$ and $\theta_{i} \in (0, 2\pi)$ for all $i$. For a star-shaped polygon $\boldsymbol{\underline{S}}$ there exists $\boldsymbol{\theta}$ and $\boldsymbol{y}$ such that $\boldsymbol{\tilde{V}}(\boldsymbol{C}, \boldsymbol{\theta}, \boldsymbol{y}) = \boldsymbol{S}$ for any $\boldsymbol{C} \in \mathcal{K}(\boldsymbol{\underline{S}})$.
\begin{proof}
Consider the point $\boldsymbol{s}_{i} \in \boldsymbol{S}$. Since $\boldsymbol{C} \in \mathcal{K}(\boldsymbol{\underline{S}})$, there exists a line segment, $\overline{\boldsymbol{C} \boldsymbol{s}_{i}}$, from $\boldsymbol{C}$ to $\boldsymbol{s}_{i}$ that is entirely contained within the polygon $\boldsymbol{\underline{S}}$.  Let  $\theta_{i} = \text{atan2}(\boldsymbol{s}_{i, y} - C_{y}, \boldsymbol{s}_{i, x} - C_{x})$ where atan2($b$, $a$)  is the  two-quadrant arctangent representing the angle between the positive $x$-axis and the line segment from the origin to  point $(b, a)$.  Then the corresponding $\ell_{i}$ in the line set covers the line segment $\overline{\boldsymbol{C} \boldsymbol{s}_{i}}$.  By construction, the line $\ell_{i}$ intersects $\boldsymbol{s}_{i}$. The line $\ell_{i}$ also cannot intersect any other points on $\boldsymbol{\overline{S}}$, since the existence of such points would violate the assumption that $\boldsymbol{S}$ is star-shaped. Define $y_{i}  = \sqrt{(\boldsymbol{s}_{i,y} - C_{y})^{2} + (\boldsymbol{s}_{i, x} - C_{x})^{2}}$, then $\boldsymbol{s}_{i} = \boldsymbol{\hat{v}}_{i}$. Repeat this construction of $\theta_{i}$, $\ell_{i}$ and $y_{i}$ for all $\boldsymbol{s}_{i}$.  Then $\boldsymbol{s}_{i} = \boldsymbol{\hat{v}}_{i}$ for all $i$, hence $\boldsymbol{\tilde{V}}(\boldsymbol{C}, \boldsymbol{\theta}, \boldsymbol{y}) = \boldsymbol{S}$.
\end{proof}

\subsubsection{Proof of Corollary 1}
Let $\ell_{\theta}$ denote the line that extends infinitely outward from $\boldsymbol{C}$ at angle $\theta \in (0, 2\pi)$ and that intersects $\boldsymbol{\overline{S}}$. For any $\theta \in (0, 2\pi)$, the line $\ell_{\theta}$ is distinct, i.e. $\ell_{\theta} \neq \ell_{\theta'}$ for any $\theta, \theta'$ such that $\theta \neq \theta'$. \begin{proof}
Let $\theta_{i} < \theta < \theta_{i + 1}$. The line $\ell_{\theta}$ intersects $\boldsymbol{\overline{S}}$ in the line segment $\overline{\boldsymbol{s}_{i} \boldsymbol{s}_{i + 1}}$, but not at either $\boldsymbol{s}_{i}, \boldsymbol{s}_{i+1}$. Since this fact holds for any $\theta_{i}, \theta_{i+1}$, each line $\ell_{\theta}$ must be distinct.
\end{proof}

\subsection{Proof of Theorem \ref{theorem: startPoint}}
 Let  $\bold{\underline{S}}$ be a set of $N$ star-shaped polygons, let $\boldsymbol{C}$ be the true starting point, and let $\boldsymbol{\mathcal{\hat{K}}}(\bold{\underline{S}}) = \mathcal{K}(\boldsymbol{\underline{S}}_{1}) \cap \hdots \cap \mathcal{K}(\boldsymbol{\underline{S}}_{N})$ denote the intersection of the kernels of all polygons. Then, $\boldsymbol{C} \in \boldsymbol{\hat{\mathcal{K}}}(\bold{\underline{S}})$. 
\begin{proof}
 The starting point $\boldsymbol{C}$ will be in every $\mathcal{K}(\boldsymbol{\underline{S}}_{i})$ by the definition of a kernel. So, any point p that is in $\mathcal{K}(\boldsymbol{\underline{S}}_{i})$  but not $\mathcal{K}(\boldsymbol{\underline{S}}_{j})$ for any $i$, $j$ is not $\boldsymbol{C}$.  Hence, $\boldsymbol{C}$ must be in $\boldsymbol{\hat{\mathcal{K}}}(\bold{\underline{S}})$.
\end{proof}

\subsection{Proof of Theorem \ref{theorem: posArea}}

For any polygon $\boldsymbol{\underline{S}}$  that is not star-shaped  $|A_{u} (\boldsymbol{C}, \boldsymbol{\theta}, \boldsymbol{S})| > 0$ and $|A_{o} (\boldsymbol{C}, \boldsymbol{\theta}, \boldsymbol{S})|> 0$ for any $\boldsymbol{C}$ and $\boldsymbol{\theta}$. 
\begin{proof}
We show that $|A_{u} (\boldsymbol{C}, \boldsymbol{\theta}, \boldsymbol{S})| > 0 $. The proof for $|A_{o} (\boldsymbol{C}, \boldsymbol{\theta}, \boldsymbol{S})| > 0 $ is analogous.   The quantity $|A_{u} (\boldsymbol{C}, \boldsymbol{\theta}, \boldsymbol{S})|  = 0$  only if  $(\underline{\boldsymbol{S}}^{c} \cap \underline{\boldsymbol{\tilde{V}}_{u}}) = \emptyset$ and $(\underline{\boldsymbol{S}} \cap \underline{\boldsymbol{\tilde{V}}_{u}}^{c}) = \emptyset$. These sets are both empty only if $\boldsymbol{\tilde{V}}_{u} = \boldsymbol{S}$. Hence, we need show only that no $\boldsymbol{\theta}$, $\boldsymbol{C}$, $\boldsymbol{y}$ combination exists that allows $\boldsymbol{\tilde{V}}_{u} = \boldsymbol{S}$. Polygon $\boldsymbol{\underline{S}}$ is not star-shaped. So, there exists at least one point $s^{*} \in \boldsymbol{\overline{S}}$ such that for any point $\boldsymbol{C} \in \boldsymbol{\underline{S}}$, the line $\overline{\boldsymbol{C} s^{*}}$ goes outside polygon $\boldsymbol{\underline{S}}$. Since the line $\overline{\boldsymbol{C}s^{*}}$ exits $\boldsymbol{\underline{S}}$ before reaching $s^{*}$, there is at least one additional intersection point between line $\overline{\boldsymbol{C} s^{*}}$ and contour line $\boldsymbol{\overline{S}}$. Let $s^{**}$ denote this intersection.  The lines $\overline{\boldsymbol{C}s^{*}}$ and $\overline{\boldsymbol{C}s^{**}}$ are at the same angle, denoted $\theta^{*}$. For either $s^{*}$ or $s^{**}$ to be $\boldsymbol{\tilde{V}}_{u}$, we must put a a line  $\ell^{*}$ in the line set that extends at angle $\theta^{*}  = \text{atan2}(s^{*}_{y} - C_{y}, s^{*}_{x} - C_{x}) = \text{atan2}(s^{**}_{y} - C_{y},s^{**}_{x} - C_{x}) $  where atan2($b$, $a$)  is the  two-quadrant arctangent representing the angle between the positive $x$-axis and the line segment from the origin to  point $(a, b)$.  However, any corresponding selection of $y^{*}$ that results in $s^{*}$ in $\boldsymbol{\tilde{V}}_{u}$ ensures that $s^{**}$ is not in $\boldsymbol{\tilde{V}}_{u}$, since only one point in $\boldsymbol{\tilde{V}}_{u}$ is created for each  line in the line set. Hence, there is no $\boldsymbol{y}$ that would allow  $\boldsymbol{\tilde{V}}_{u} \neq \boldsymbol{S}$, so $|A_{u} (\boldsymbol{C}, \boldsymbol{\theta}, \boldsymbol{S})| > 0$.
\end{proof}

\section{Appendix: Computation of {$\hat{\boldsymbol{\mathcal{K}}}({\bold{S}}$)}}
\label{app: compK}
Computation of $\boldsymbol{\hat{\mathcal{K}}}(\bold{\underline{S}})$ is simple.  Each $\mathcal{K}(\boldsymbol{\underline{S}}_{i})$  is  the intersection of a set of interior half-planes with one half-plane defined by each edge $\boldsymbol{\overline{S}}_{i}$. For each edge, the plane is divided with the line that intersects the edge. The half-plane that is on the same side of the dividing lines as the interior of the polygon is used.  The kernel of $\boldsymbol{S}_{i}$,  $\mathcal{K}(\boldsymbol{\underline{S}}_{i})$,  is the intersection of all these interior half-planes \citep{Shamos1975}. Intersecting all the individual polygons, $\mathcal{K}(\boldsymbol{\underline{S}}_{i})$, produces $\boldsymbol{\hat{\mathcal{K}}}(\bold{\underline{S}})$. For any $\boldsymbol{\underline{S}}_{i}$ with $n$ edges, $\mathcal{K}(\boldsymbol{\underline{S}}_{i})$ can be found as described in $O(n \log n)$ time \citep{Shamos1975}. An alternative algorithm can find $\mathcal{K}(\boldsymbol{\underline{S}}_{i})$  in $O(n)$ time \citep{Lee1979}.

\section{Supplement: Simulation shape parameters}
\label{supp: simPars}
This section gives a list of  the parameter values used in simulations. All values are rounded to the nearest thousandth. For \textit{Shape A}, \textit{Shape B}, and \textit{Shape C}, all the following parameters are shared.
\begin{align*}
\kappa = 2 & \\
\boldsymbol{C} = (&0.5, 0.5)\\
\nonumber \boldsymbol{\theta}  = (&0.063, 0.188, 0.314, 0.440, 0.565, 0.691, 0.817,
 0.942, 1.068, 1.194, 1.319, 1.445, 1.571, 1.696, 1.822, \\ 
&1.948, 2.073, 2.199, 2.325, 2.450, 2.576,  2.702, 2.827, 2.953, 3.079, 
 3.204, 3.330, 3.456,   3.581, 3.707, \\ 
 & 3.833, 3.958, 4.084, 4.210, 4.335,
 4.461, 4.587, 4.712, 4.838, 4.964, 5.089, 5.215,  5.341, 5.466, 5.592,   \\ 
 &5.718, 5.843, 5.969, 6.095, 6.220)\\
\nonumber \boldsymbol{\sigma} =  (&0.035, 0.044, 0.053, 0.062, 0.071, 0.080, 0.080,
0.073, 0.065, 0.058, 0.050, 0.042, 0.035, 0.035, 0.035,  \\ 
 &0.035, 0.035, 0.035, 0.035, 0.035, 0.035,
 0.035, 0.035, 0.035, 0.035, 0.035, 0.044, 0.053,
 0.062, 0.071,\\ 
 &0.080, 0.080, 0.073, 0.065, 0.058, 0.050, 0.042, 0.035, 0.035, 0.035, 0.035, 0.035,
 0.035, 0.035, 0.035,\\
 &  0.035, 0.035, 0.035, 0.035,0.035)
 \end{align*}

\noindent The mean parameters  for \textit{Shape A}, \textit{Shape B}, and \textit{Shape C} are 
\begin{align*}
\nonumber \boldsymbol{\mu}_{A}  = (&0.294, 0.304, 0.306, 0.290, 0.264, 0.241, 0.220,
0.213, 0.219, 0.239, 0.259, 0.282, 0.298, 0.300, 0.276, \\ 
& 0.254, 0.236, 0.216, 0.211, 0.217, 0.212,
0.206, 0.200, 0.192, 0.195, 0.239, 0.287, 0.313,
 0.318, 0.321, \\ 
& 0.322, 0.321, 0.319, 0.316, 0.312, 0.308, 0.280, 0.240, 0.197, 0.197, 0.214, 0.231,
0.247, 0.264, 0.263, \\
&0.256, 0.249, 0.244, 0.270, 0.283)\\
\nonumber \boldsymbol{\mu}_{B}  =  (&0.3, 0.3, 0.3, 0.3, 0.3, 0.3, 0.3, 0.3, 0.3, 0.3, 0.3,
0.3, 0.3, 0.3, 0.3, 0.3, 0.3, 0.3, 0.3, 0.3, 0.3, 0.3,  0.3, 0.3,\\ 
 &
  0.3, 0.3, 0.3, 0.3, 0.3, 0.3, 0.3, 0.3, 0.3, 0.3, 0.3, 0.3, 0.3, 0.3, 0.3, 0.3, 0.3, 0.3, 0.3, 0.3,
0.3, 0.3, 0.3, 0.3,\\
& 0.3, 0.3)\\
\nonumber \boldsymbol{\mu}_{C} = (&0.300, 0.263, 0.225, 0.188, 0.150, 0.150, 0.187,
  0.225, 0.262, 0.300, 0.290, 0.243, 0.197, 0.150, 0.150,  \\
 &0.200, 0.250, 0.300, 0.300, 0.274, 0.248,
 0.222, 0.196, 0.170, 0.170, 0.200, 0.230, 0.260,
0.290, 0.320,  \\ 
&0.320, 0.298, 0.275, 0.253, 0.230,
 0.150, 0.200, 0.250, 0.300, 0.350, 0.360, 0.312,
 0.265, 0.218, 0.170,   \\ 
 &0.170, 0.203, 0.235, 0.268,
 0.300).
 \end{align*}

\end{document}